# Ultrafast studies on the photophysics of matrix-isolated radical cations of polycyclic aromatic hydrocarbons. [1]

Liang Zhao,[b] Rui Lian,[a] Ilya A. Shkrob,[a] Robert A. Crowell,[*,a] Stanislas Pommeret,[a,c] Eric L. Chronister,[b] An Dong Liu,[a,2] and Alexander D. Trifunac[a]

[a] *Chemistry Division, Argonne National Laboratory, Argonne, IL 60439*
[b] *Chemistry Department, University of California at Riverside, Riverside, CA 92521*
[c] *CEA/Saclay, DSM/DRECAM/SCM/URA 331 CNRS 91191 Gif-Sur-Yvette Cedex, France*

**ABSTRACT**

Ultrafast relaxation dynamics for photoexcited PAH cations isolated in boric acid glass have been studied using femtosecond and picosecond transient grating spectroscopy. With the exception of perylene[+], the recovery kinetics for the ground doublet ($D_0$) states of these radical cations are biexponential, containing a fast (< 200 fs) and a slow (3-20 ps) components. No temperature dependence or isotope effect was observed for the fast component, whereas the slow component exhibits both the H/D isotope effect (1.1-1.3) and strong temperature dependence (15 to 300 K). We suggest that the fast component is due to internal $D_n$ to $D_0$ conversion and the slow component is due to vibrational energy transfer (VET) from a hot $D_0$ state to the glass matrix. The observed rapid, efficient deactivation of the photoexcited PAH cations accounts for their remarkable photostability and have important implications for astrochemistry, as these cations are the leading candidates for the species responsible for the diffuse interstellar bands (DIB) observed throughout the Galaxy.

---

[2] Permanent address: Institute of Low Energy Nuclear Physics, Beijing Normal University, Beijing, China.

* To whom correspondence should be addressed: *Tel* 630-2528089, *FAX* 630-2524993, *e-mail:* rob_crowell@anl.gov.

# 1. Introduction

Processes involving radical cations of polycyclic aromatic hydrocarbons (PAHs) occur in many areas of chemistry, such as organic synthesis, photo- and radiation chemistry, and astrochemistry. In particular, these radical cations are thought to be responsible for diffuse interstellar bands (DIBs) in star-forming clouds in our Galaxy.[1-4] While the association of these ubiquitous bands with the PAH cations is still tentative,[1] these radical cations, together with their parent molecules and corresponding carbonium ions, could be the prevalent form of the organic matter in the Universe.[4] Why are these exotic - by terrestrial standards - species so common? Some researchers speculate that the PAHs are generated in mass-losing carbon stars, in ion-molecule reactions of neutral and ionized C atoms, by condensation of C chains, or by thermo- and photo- induced condensation of molecules adsorbed on dust particles (see, for example, ref. 4). In these scenarios, the high abundance of the PAH cations is due to high production rate of their parent molecules. Other researchers focus[1,3,5] on the dynamics of the gas-phase carbonaceous molecules and ions in the interstellar medium where these species are exposed to intense radiation from young stars. It seems that PAH cations of intermediate size are favored in this harsh environment.[3,5] Regardless of the exact answer, it is certain that (i) PAH cations are products of lengthy chemical evolution driven by heat and radiation, and (ii) these cations are abundant in space because they are more photostable than most neutral molecules and anions.[1-4] In particular, the primary decay processes for photoexcited radical cations of PAHs are nonradiative:[5-10] almost all of the excitation energy is converted into heat that (in the interstellar medium) is emitted as IR radiation.[1,2] It is the purpose of this work to study the photophysical processes of PAHs.

While the photophysics and reactivity of the excited singlet and triplet states of neutral PAH molecules is well understood, very little is known about the photophysics of their radical cations. In the gas phase, -H, -H$_2$, and -C$_2$H$_2$ photofragmentation of C$_{10}$-C$_{16}$ PAH cations have been studied by mass-spectrometry.[5] Typical dissociation rates for 7 eV excess energy are $(1-3) \times 10^3$ s$^{-1}$, and the onsets of dissociation are 4-4.5 eV (which is close to the C-H bond dissociation energy).[5] These gas-phase studies demonstrate remarkable photostability of PAH cations (as compared to their parent molecules) and suggest fast energy relaxation in these species. There have also been numerous EPR, UV-VIS, and IR studies of matrix-isolated PAH cations. From these studies and concurrent *ab initio* and density functional theory calculations, a wealth of data on the structure and energetics of aromatic radical cations has emerged (e.g., see Table 1 and references given therein).

Recently, Vauthey and coworkers[7,8] examined the relaxation dynamics of photoexcited matrix-isolated radical cations of several organic molecules, including perylene$^+$ and tetracene$^+$. The 640 nm (D$_0 \leftarrow$ D$_1$) band in the fluorescence spectrum of perylene$^+$ was observed, and a quantum yield of 10$^{-6}$ obtained. Picosecond (for perylene$^+$) and subpicosecond (for tetracene$^+$) transient grating (TG) spectroscopy was used to observe the recovery of the D$_0$ state following laser photoexcitation (in the D$_0 \rightarrow$ D$_5$ and D$_0 \rightarrow$ D$_1$ bands, respectively). This study brought an unexpected result: although the D$_1 \rightarrow$ D$_0$ conversion in tetracene$^+$ and perylene$^+$ was fast (25 to 100 ps, depending on the matrix)[7,8] as compared to the typical times of 10$^{-9}$-10$^{-6}$ s for S$_1 \rightarrow$ S$_0$ transition in neutral PAH molecules,[9] this conversion was much slower than the typical times of 10-500 fs for nonradiative $S_n \to S_{n-1}$ transitions that involve higher excited singlet states ($S_n$) of these PAH molecules.[9,10] Since the rate of nonradiative transition broadly correlates with the energy gap between the initial and final states,[10] these energetic $S_n$ states should have provided a good reference system for the lower doublet states of aromatic radical cations, as the corresponding energy gaps are comparable (0.5-1 eV; see Table 1). Since tetracene$^+$ and perylene$^+$ have unusually large $D_1$-$D_0$ gaps of 1.44 and 1.56 eV,[8] respectively, it appears that these two radical cations might represent the exception rather than the rule. To observe nonradiative transitions in a typical PAH cation, time resolution better than 1 ps is needed.

In this work, photophysics of radical cations of perprotio and perdeuterio anthracene, naphthalene, biphenyl, and perylene stabilized in boric acid glass are studied using femtosecond transient grating spectroscopy. Our results suggest that for most of the PAH cations, the excited electronic states relax through a nonradiative $D_n \to D_0$ transition that occurs in less than 200 fs. This process rapidly converts the electronic energy into vibrational energy of the ground $D_0$ state which then undergoes vibrational energy transfer (VET), on a picosecond time scale, by heat transfer to the matrix. The typical VET times observed in the boric acid glass are 5 to 20 ps. Perylene$^+$ has an exceptionally long lifetime for the $D_1 \to D_0$ conversion, ca. 19 ps, by virtue of its unusual energetics. To save space, some data are given in the Supporting Information. Figures and tables with a designator "S" after the number (e.g., Fig. 1S) are placed therein.

## 2. Experimental

*Sample Preparation.* Orthoboric acid ($H_3BO_3$), biphenyl-$h_{10}$ and -$d_{10}$, naphthalene-$h_8$ and -$d_8$, anthracene -$h_{10}$ and -$d_{10}$, pyrene-$h_{10}$, and perylene-$h_{12}$ (see Fig. 1S for the structure) of the highest purity available from Sigma-Aldrich were used as received. PAH-doped glass was prepared by adding crystalline PAHs to the boric acid melt at 200-240°C. The melt was cast between two thin windows made of fused silica, calcium fluoride, or sapphire and produced high quality, optically clear glasses upon cooling. Typical glass film thickness was 100 to 400 μm. These samples were then exposed to 5 to 50 pulses of 248 nm light (15 ns fwhm, 0.05 J/cm$^2$) from a Lambda Physik model LPx-120i KrF excimer laser, at room temperature. Absorption spectra [11,12] confirmed the formation of radical cations [11] with conversion efficiency better than 80%. The concentrations of the radical cations were spectrophotometrically determined to be 0.1-0.5 mM (for perylene$^+$, ca. 20 μM). Because of a large nonresonant TG signal observed in the windows, the glass film was removed from the substrate and transparent 2 mm x 2 mm pieces were mounted on the cold finger of a closed cycle helium refrigerator (Lake Shore Cryogenics) operable from 10 to 300 K. [13] Typical absorbance of these films at the excitation wavelength was 0.3-1. For absorbances < 0.1, the nonresonant TG signal from the glass matrix was superimposed on the TG signal from PAH cations.

*Transient grating spectroscopy.* The ground state recovery dynamics of PAH cations were studied using femtosecond TG spectroscopy. Details of this technique can be found elsewhere. [7,8,14] Briefly, a standard three beam transient grating setup using a folded BOXCARS geometry with crossing angles of ca. 2° was used. [7,8,14] Calcite Glan-Thomson

6.

polarizers were placed in each of the three beam paths and the scattered probe beam to ensure pure and parallel polarization so that the observed signal reflected the dynamics of $\chi^{(3)}_{1111}$ element of the third-order nonlinear susceptibility tensor.[7,14] Under the conditions of our experiments a density grating (due to heat-induced expansion of the sample) is produced simultaneously to the population grating that we are interested in observing. The density grating generates an acoustic standing wave that over time will affect the population grating. The density grating has no effect on our measurements because of the small angle geometry used in this set-up (~2°) results in an acoustic period of >10 ns. The instrument response was determined by obtaining a nonresonant TG signal due to optical field induced electronic and nuclear Kerr effect in a thin plate of fused silica or undoped boric acid glass, before and after each measurement. This response was Gaussian, with a typical fwhm of 60-70 fs (for femtosecond kinetics) or 160 fs (for picosecond kinetics).

The tunable source of femtosecond light pulses used in the experiments is an optical parametric amplifier (Spectra Physics OPA model 8000CF). The OPA is pumped by a 800 µJ, 800 nm, 55 fs fwhm pulse split off from a two-stage Ti:Sapphire based chirped pulse amplifier system (3.5 mJ) operated at a repetition rate of 1 kHz. Passing the output of the OPA through a pair of SF-10 prisms resulted in transform limited pulses of 45-55 fs duration (depending on the wavelength). The pump energy (in each pump beam) was less than 500 nJ and the probe energy was less than 50 nJ. To subtract the background signal, one of the pump beams was chopped at 500 Hz, and the photodiode signal fed into a digital lock-in amplifier (SRS model 810). Otherwise, the detection electronics were identical to those described previously.[15] To obtain subpicosecond kinetics, 50-100 traces acquired with a time constant of 30 ms were averaged. The delay times of the probe pulses were changed either linearly ($\Delta t$=15 fs) or on a quasilogarithmic grid (Fig. 1). [15] The error bars shown in some kinetics are 95% confidence limits. To obtain picosecond kinetics, the delay time of a 160 fs pulse was changed in steps of 150-200 fs and 20-30 traces acquired with a time constant of 300 ms were averaged.

3. Results

Fig. 1 shows a TG signal observed for the radical cation of naphthalene-$h_8$ in a UV-irradiated boric acid glass at 300 K. Only naphthalene$^+$ absorbs VIS light in this sample. The kinetics were obtained using 680 nm pump and probe pulses of 63 fs fwhm. The pump energy (1.82 eV) corresponds to the excitation of a strongly allowed $D_0 \rightarrow D_2$ transition (Fig. 2S and Table 1). Following the excitation pulse, the recovery of the photobleached $D_0$ state is observed. The intensity of the grating signal is proportional to the square of the concentration of photoexcited cations. For an exponential process, the TG signal decays at twice the rate of the $D_0$ state recovery. Note that only population grating [7,14] contributes to the TG signal observed at the end of the pump pulse. During photoexcitation, a nonresonant signal due to the optical Kerr effect (Fig. 1) [8,14] contributes to the TG signal. This signal makes it difficult to observe rapid processes that occur in less than 50 fs.

Following the 680 nm photoexcitation of naphthalene$^+$, there is a short-lived TG signal whose single-exponential decay kinetics corresponds to a lifetime $\tau_f$ of 195±5 fs (Figs. 1 and 3S(a)). After the decay of this short-lived signal, there is a slower TG signal that decays over the time period of tens of picoseconds (Figs. 1, 2(b), and 3S(b)). This slow signal comprises ca. 10% of the initial "spike" observed at short delay times (the apparent relative weight of the slow component increases for longer excitation pulses). Using a longer (160 fs) pulse, we were able to obtain a better quality decay kinetics for t > 1 ps that are shown separately in Fig. 2(b), for naphthalene-$d_8^+$. For all PAH cations, the slow kinetics can be fit using a single exponential function with a time constant $\tau_s$ of a few picoseconds (e.g., Figs. 1 and 2(b)). Both of the time constants, $\tau_f$ and $\tau_s$, showed no variation with the concentration of the aromatic dopant, the extent of ionization by the 248 nm light, or the amount of water in the boric acid glass.

The overall kinetics (trace (i) in Fig. 1) can be simulated by convolution of the Gaussian response function (trace (iii) in Fig. 1) with a biexponential function. Figs. 1 and 3S(a) demonstrate a comparison between the kinetics obtained for radical cations of naphthalene-$h_8$ and -$d_8$. While the time constant of the fast decay does not change with H/D substitution (Fig. 1, trace (i) and (ii) and Fig. 3S(a)), the decay of the slow component for the perdeuterio cation is 1.24 times faster (Table 2). Cooling the sample also changes this slow component (Fig. 2). For naphthalene-$d_8^+$, $\tau_s$ becomes progressively longer as the temperature decreases (Table 2) and the relative weight of the slow component increases by 50% upon cooling from 300 to 70 K (Fig. 2(b)). Unlike the fast component, the slow decay kinetics are sensitive both to the sample temperature and H/D substitution in the aromatic radical cation.

Similar results were obtained for some other photoexcited PAH cations, Figs. 4S, 5S, and 6S (the energetics are given in Table 1 and the life times are given in Table 2). An increase in $\tau_s$ upon cooling from 295 K to 15 K was observed for naphthalene-$h_8^+$ (increase of 2.8 times), biphenyl-$h_{10}^+$ (increase of 1.47 times, Fig. 6S(a)), biphenyl-$d_{10}^+$ (1.25 times, Fig. 6S(b)), and anthracene-$d_{10}^+$ (1.27 times, Fig. 5S(b)). The only two exceptions are perylene-$h_{12}^+$ (Fig. 7S) and anthracene-$h_{10}^+$ (Fig. 5S(b)) for which the change in $\tau_s$ with the temperature was negligible (Table 2). For anthracene$^+$, the relative weight of the slow component increased and $\tau_s$ decreased upon H/D substitution (Figs. 4S(a) and 5S(b)). The decrease in $\tau_s$ was higher at 295 K than at 15 K. For example, for anthracene$^+$ the kinetic isotope effect $\alpha_{H/D}$ defined as the ratio $\tau_s(h_{10})/\tau_s(d_{10})$ is 1.32±0.09 at 295 K and 1.03±0.04 at 15 K. For biphenyl$^+$, the isotope effect is reversed as compared to anthracene$^+$ and naphthalene$^+$: Biphenyl-$d_{10}^+$ exhibits *longer* $\tau_s$ than biphenyl-$h_{10}^+$ and the slow component has *higher* relative weight in the perdeuterio cation (Fig. 6S). Thus, only for biphenyl$^+$, does the decay rate of the slow component obey Siebrand's rule [10] for radiationless electronic transitions (which posits slower rates for D-substituted molecules with $\alpha_{H/D}$ of 0.75). As argued in the Discussion, this is due to the fact that the slow component is due to *vibrational* rather than *electronic* relaxation.

For anthracene -$h_{10}^+$ and -$d_{10}^+$, the fast component decays faster ( < 50 fs) than the time resolution of our TG spectrometer, and the time profile of the initial "spike" is close to that of the pump pulse (Figs. 4S(a) and 5S(a)). For biphenyl -$h_{10}^+$ and -$d_{10}^+$, the fast component is clearly observed (Fig. 4S(b)), yielding $\tau_f$=118±8 fs. As in naphthalene$^+$, the time constant $\tau_f$ of the fast component does not change upon the H/D substitution (Fig. 4). For these cations, the relative weight of the slow component correlates inversely with $\tau_f$, increasing from anthracene$^+$ to biphenyl$^+$ to naphthalene$^+$ (Figs. 1, 4S, and 5S(a)).

Perylene$^+$ is different from other radical cations that we studied: the subpicosecond component is lacking and the time constant of the slow component does not change from 15 K to 295 K (Fig. 7S). All picosecond kinetics are monoexponential and can be fit with a temperature-independent $\tau_s$ of ca. 18.8 ps (Figs. 2(a) and 7S). This time is notably shorter than a value of 35±3 ps obtained by Gumy and Vauthey [7] from a picosecond TG experiment on the same photosystem (532 nm, 25 ps fwhm pulse excitation). Our measurement is perhaps more accurate owing to the faster time resolution of our setup. Brodard et al. [8] reported a considerable temperature and isotope effect ($\alpha_{H/D}$ of ca. 1.23 at 295 K) on the decay rate of TG signal for perylene$^+$ in sulfuric acid. Activation energies of 40 to 80 meV (300 to 350 K) that varied with the isotope composition of the dopant and the substrate were obtained (Table 2 in ref. 8). We did not observe a temperature effect for perylene-h$_{12}$$^+$ in the boric acid glass (Fig. 2).



## 4. Discussion

In our experiments, two relaxation regimes for the recovery kinetics of the $D_0$ state are observed: a subpicosecond regime and a picosecond regime. Given that these cations are photoexcited to their $D_2$, $D_3$, or even $D_5$ states, it is tempting to associate the fast component with a rapid (possibly, stepwise) $D_n \rightarrow D_1$ nonradiative transition and the slow component with a slower (also nonradiative) $D_1 \rightarrow D_0$ transition - in full analogy with the electronic deactivation in neutral PAH molecules.

For aromatic singlets, the slowest (usually, radiative) transition is from their first excited $S_1$ states to their ground $S_0$ states (Kasha's rule) whereas less energetic $S_n \rightarrow S_{n-1}$ transitions are nonradiative and occur much faster (< 10 ps). [9,10] Typically, the energy gaps between the $S_n$ and $S_{n-1}$ states are 2,000 to 6,000 cm$^{-1}$ (vs. 10,000-25,000 cm$^{-1}$ for the $S_1$ and $S_0$ states) and the transition times are 10-20 fs to 1-5 ps (vs. 1 ns to 1 μs for the $S_1$ and $S_0$ states). [9] For the doublet manifold of PAH cations, the energy gaps between the $D_0$ and $D_1$ states are 7,000 to 10,000 cm$^{-1}$ (see Table 1) and one would expect to observe rapid nonradiative transitions in the doublet manifold. Though an approximate correlation between the energy gap and the corresponding relaxation rate exists (the energy gap law, see chapter 2.11 in ref. 10), these rates can vary strongly between different photosystems, even if the energetics are similar (see, for example, Table 4 in ref. 9). Generally, the energy gap law is valid for relatively slow nonradiative transitions (such as $T_1 \rightarrow S_0$ and $S_1 \rightarrow T_1$ transitions); [10] for *rapid* transitions (such as $S_n \rightarrow S_{n-1}$ transitions), the law was shown to have limited applicability. [9]



Cursory examination of Tables 1 and 2 and Figs. 2S and 8S suggests that $\tau_f$ and $\tau_s$ times obtained at 295 K do correlate with the $D_n$-$D_1$ and $D_1$-$D_0$ energy gaps, respectively (though no such correlation is apparent in the 15 K data). However, detailed examination of the kinetics given in Appendix 1 in the Supporting Information indicates that the stepwise electronic relaxation is not supported by our results.

We suggest that the subpicosecond component observed for all PAH cations with exception of perylene$^+$ corresponds to rapid internal $D_n \to D_0$ conversion in the doublet manifold, whereas the picosecond decay is due to VET from a hot $D_0$ state formed in the course of this rapid electronic relaxation (Fig. 3). Such an interpretation readily accounts for the lack of the isotope and temperature effects on the decay kinetics of the fast component. Since the rate of VET to the glass matrix depends on the matrix temperature and available vibrational modes in the donor, $\tau_s$ changes both with sample cooling and H/D substitution in the PAH cation (see below). The decrease in the relative weight of the slow component at cryogenic temperatures is also rationalized: In a hot $D_0$ state, the spectral line is broader than the same line in a thermalized cation.[20a] For a given probe wavelength, the difference between the absorbances of the hot and relaxed $D_0$ states strongly depends on the overlap between the spectra of these two states: the greater is the overlap, the smaller is the slow TG signal. At the lower temperature, the line is narrower, the spectral overlap becomes worse, and the weight of the slow TG signal increases.

In perylene$^+$, due to an exceptionally large $D_0$-$D_1$ gap, the internal conversion is inhibited, and the 18.8 ps decay kinetics reflects the slow internal conversion of the $D_1$

13.

state to the ground $D_0$ state. Since for perylene$^+$ (and, possibly, tetracene$^+$) [8] the slow component is due to electronic rather than vibrational relaxation, the contrast between temperature dependencies shown in Fig. 2(a) is accounted for. Though our data are insufficient to establish the details of internal conversion in the doublet manifold, we believe that in all cases considered (with the exception of perylene$^+$), the fast component is from the $D_1 \rightarrow D_0$ transition; the relaxation of the higher $D_n$ states to the $D_1$ state is so fast that it cannot be observed with our instrumentation.

Recently, Nishi and co-workers [21] studied vibrational relaxation of naphthalene$^+$ and biphenyl$^+$ cations generated in biphotonic ionization of their parent molecules in polar liquids (267 nm, 4 ps fwhm pulses were used for photoexcitation). Transient kinetics for Raman bands corresponding to inter-ring C-C stretching modes were obtained. Both the Raman band centers and widths change on the picosecond time scale. For biphenyl$^+$ in 1-butanol, ethyl acetate, and acetonitrile, time constants of < 5, 13, and 17 ps, respectively, for thermalization of the C-C stretching bands were obtained. For naphthalene$^+$ and *trans*-stilbene$^+$ in acetonitrile, these time constants were 20-30 and 40 ps, respectively. Häupl et al. [22] studied femtosecond dynamics of photoexcited ($D_1$) radical cation of methylviologen in acetonitrile and concluded that the electronic relaxation of the $D_1$ state occurred in < 700 fs (<350 fs in water [23]). The resulting vibrationally excited $D_0$ state with lifetime of 16 ps (2 ps in water [23]) had the same absorption spectrum as the parent cation but was shifted by 810 cm$^{-1}$ to the red. This shift is equivalent to the frequency of the intra-ring C-C stretch. In addition to this "hot" $D_0$ state, there was another vibrationally excited $D_0$ state with lifetime of 1 ps that was

14.

populated with the same rate as the long-lived state. These several examples indicate that VET from PAH cations to the polar solvent typically occurs over 1-50 ps. Similar rates of VET to liquid solvent and solid host matrices (vibrational cooling) were observed for vibrationally-excited PAH molecules in their $S_0$ and $S_1$ states. [20b]

The slow components observed in our TG experiments for naphthalene$^+$, biphenyl$^+$, and anthracene$^+$ have time constants between 5-10 ps (295 K) and 10-25 ps (15 K). These lifetimes are somewhat shorter than the typical lifetimes for vibrationally excited $D_0$ states in polar solvents (10-40 ps) [21,22]. VET in a rigid matrix, such as boric acid glass, could be faster than in an organic liquid, because the thermal conductivity of the glass is much higher (e.g., at 300 K, the thermal conductivity of vitreous boron trioxide is 3 times that of ethyl alcohol). [24] Besides, the initial stage of VET to the glass matrix may involve other than the high-frequency C-C stretching modes observed in the Raman experiments. [21] Nishi and co-workers suggest that the VET from a hot PAH cation to the matrix occurs not in one but two stages. The relatively slow rate of heat transfer observed in the time-resolved Raman experiments corresponds to cooling of the first solvation shell around the radical cation; this process is preceded by a faster VET from a hot radical cation to the first solvation shell. Due to insufficient time resolution, this fast process was not observed in the Raman experiments. This two-step mechanism has also been invoked to explain the vibrational relaxation of PAH molecules in transient absorption, grating, and Raman experiments of Sukowski et al. and others. [25]

The two-step mechanism qualitatively accounts for the isotope and temperature dependencies obtained in our TG experiments. The H/D isotope effect may originate

15.

through a VET that involves low-frequency C-H modes of a PAH cation [26] and vibrational modes of two glass-forming $B_3O_6$ rings that sandwich this planar cation between them (see Appendix 2 in the Supporting Information). Such a process should not depend on the sample temperature. By contrast, cooling of the first solvation shell depends on the sample temperature whereas it does not depend on the isotope composition of the dopant. In the two-step model, the overall kinetics for the picosecond component can be explained in the following way: When the temperature is low, the decay rate for the tail in the TG kinetics is limited by the (slow) rate of heat transfer to the glass bulk, and the isotope effect is negligible. At room temperature, the rate of the heat transfer from the first solvation shell is comparable to the rate of VET from the hot $D_0$ state of the radical cation to the first solvation shell, and the decay kinetics of the TG signal exhibit isotope effects.

*Implications for photochemistry.* While PAH cations are photostable both in the gas phase and in the host media whose molecules have high ionization potentials (such as acetonitrile), in most matrices a hole injection [27,28] occurs: photoexcited radical cation ($Ar^+$) oxidizes the solvent. This reaction occurs both in nonpolar liquids, such as saturated hydrocarbons (where the resulting solvent radical cations is metastable, reaction (1)) [27] and in protic polar solvents (where the photoinduced charge transfer is concerted with a proton transfer, reaction (2)). [28]

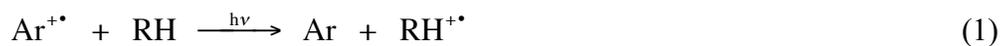

$$Ar^{+\bullet} + RH \xrightarrow{h\nu} Ar + RH^{+\bullet} \qquad (1)$$

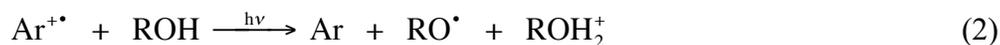

$$Ar^{+\bullet} + ROH \xrightarrow{h\nu} Ar + RO^{\bullet} + ROH_2^{+} \qquad (2)$$

16.

The typical quantum yields for these two reactions are given in Table 1S in the Supporting Information. Since the heats of reactions (1) and (2) depend on the difference between the ionization potentials (IP's) of the parent aromatic molecule and the solvent, these quantum yields correlate with the molecular IP (Table 1S and refs. 27 and 28). Interestingly, some PAH cations, such as anthracene$^+$ and biphenylene$^+$, exhibit very low yields of hole injection, though the corresponding IP's are sufficiently high and the absorbance of the cation at the excitation wavelength is strong. [28] Conversely, photoexcited perylene$^+$ oxidizes both polar and nonpolar solvents, though the corresponding reactions are barely exothermic. [27,28] Our kinetic data rationalize these observations. Photoexcited anthracene$^+$ is extremely short-lived, and the hole injection involving this species is inefficient. Photoexcited perylene$^+$ is very long-lived, and even slow, inefficient hole injection can occur. From the data of Tables 1S and 2, we estimate that for naphthalene$^+$ and biphenyl$^+$ in 2-propanol, reaction (2) occurs with rate constant of $(3-5) \times 10^{12}$ s$^{-1}$, whereas for perylene, this rate constant is just $10^8$ s$^{-1}$. Note that if the lifetimes of the electronically-excited radical cations involved in the "hole injection" were given by $\tau_s$ (as in the first model considered above) rather than by $\tau_f$ (as in the second model) the anomalous behavior of anthracene$^+$ would be difficult to account for, as this cation has longer $\tau_s$ than naphthalene$^+$ and biphenyl$^+$ (Table 2).

*Implications for astrochemistry.* Recently, Snow, Zukowski, and Massey [32] have undertook an extensive study of the region surrounding the strong DIB line centered at 442.8 nm, by surveying 35 young stars located in Cyg OB2 association. They have found that the profile of this line is symmetric around the center, invariant across the sky, and



can be well fit with a Lorentzian line whose width corresponds to the upper state lifetime of 380 fs. Their estimate is within the same order of the magnitude as the time constants for rapid recovery of the $D_0$ states observed in our matrix isolation experiments. Thus, our observations strengthen the link between the DIB lines and PAH cations. In another recent study, Biennier, Salama, Allamandola, and Scherer [33] have used pulsed discharge nozzle cavity ringdown spectroscopy to study the photophysics of cold, gas-phase PAH cations. For the same $D_2$ band of naphthalene$^+$ studied in this work, they obtained the lifetime of 210 fs (vs. 190±5 fs obtained in our experiments). This result suggests that the lifetimes of electronically excited states in the gas phase and matrix isolated PAH cations are comparable.

**6. Concluding Remarks.**

Femto- and pico- second transient grating kinetics for matrix-isolated photoexcited PAH cations can be understood in terms of two processes (Fig. 3): (i) a very rapid (< 200 fs) internal conversion to a hot $D_0$ state and (ii) a vibrational energy transfer from this state to the solid matrix (4-25 ps). A two-step mechanism [21] for VET from the hot $D_0$ state to the glass matrix is suggested. The ultrafast internal conversion accounts for the remarkable stability of PAH cations towards VIS and near UV photoexcitation.

Given the high rates of electronic deactivation and hole injection for aromatic radical cations, we conclude that (with few exceptions) a participation of electronically-excited PAH cations in exothermic electron-transfer reactions [29] is unlikely. Even if hole injection [27,28] does not occur, the excited PAH cations rapidly deactivate in less than 1 ps. On the other hand, both our TG results and other recent studies [21,22,23] suggest that

18.

vibrational cooling of organic radical cations is relatively slow. This cooling can occur on the same time scale as charge recombination, and the formation of vibrationally excited cations - whose reaction properties can be different from the ground state cations - should be taken into account.

## 7. Acknowledgment


IAS thanks Dr. F. Salama and Profs. T. P. Snow, T. Oka, and V. M. Bierbaum for helpful discussions and Dr. M. C. Sauer, Jr. for technical assistance. This work was performed under the auspices of the Office of Basic Energy Sciences, Division of Chemical Science, US-DOE under contract No. W-31-109-ENG-38. SP acknowledges the support of the DGA through the contract number DSP/01-60-056.


**Supporting Information Available:** (1.) Appendix 1: two-step electronic relaxation scenario rebuffed; (2.) Appendix 2: intermolecular VET in boric acid glass; (3.) Additional references; (4.) Table 1S: Quantum yields for 532 nm photon induced "hole injection" in the room-temperature *trans*-decalin and 2-propanol; (5.) Captions to Figs. 1S to 8S; (6.) A PDF file containing Figs. 1S to 8S. This material is available free of charge via the Internet at http://pubs.acs.org.



**References.**

**Table 1.**

Energetics of the PAH cations.

| Aromatic Cation | $D_0$ [a] | Band Excited [b] | Energy of Band Excited nm (eV) | Symmetry of State Excited. | $D_1$ | $E(D_1)-E(D_0)$, eV [b] | $E(D_1)-E(D_0)$ eV [c] |
|---|---|---|---|---|---|---|---|
| biphenyl$^+$ | $^2B_{2g}$ | $D_3$ | 680 (1.82) | $^2B_{3u}$ | $^2A_u$ | 0.7 (ref. 17b) 0.88 (ref. 17a) | 1.53 (0.0062) |
| naphthalene$^+$ | $^2A_u$ | $D_2$ | 680 (1.82) | $^2B_{2g}$ | $^2B_{1u}$ | 0.78 (ref. 17a) 0.72 (ref. 30) | 0.894 [16] |
| anthracene$^+$ | $^2B_{2g}$ | $D_2$ | 720 (1.72) | $^2A_u$ | $^2B_{3g}$ | 1.13 (ref. 30) | 1.33 [16] |
| pyrene$^+$ | $^2B_{3g}$ | $D_5$ |  | $^2A_u$ | $^2B_{2g}$ | 0.85 (ref. 29) | 0.85 [29] 1.17 [16] |
| perylene$^+$ | $^2A_u$ | $D_5$ | 540 (2.3) | $^2B_{3g}$ | $^2B_{3g}$ | 1.56 (ref. 16) [d] | 1.596 (0.0068) [16] |

a) the ground state, assuming $D_{2h}$ symmetry; for ground-state biphenyl$^+$, $D_2$ symmetry with torsion angle ≈ 20-40° between the aromatic rings is more appropriate (ref. 17).
b) relative energy of the first excited state as determined by photoelectron spectroscopy.
c) the same, by theoretical calculation (in parenthesis, the oscillator strength for the allowed $D_0 \rightarrow D_1$ transitions is indicated);
d) the same, from the onset of the *0-0*, $D_0 \rightarrow D_1$ transition.

Computation methods are LNDO/S + PERT CI (ref. 17e), HMO (ref. 17b), QCFF/PI + CI (ref. 16), and CIPSI/PPP and CIPSI/PI + CI (ref. 29)



**Table 2**

Kinetic parameters for PAH cations in boric acid glass.

| aromatic cation | Band Excited | $\tau_s$ at 295 K, ps [a] | $\tau_s$ at 15 K, ps [a] | $\tau_f$, (fs) [b] |
|---|---|---|---|---|
| biphenyl$^+$ | D$_3$ | 4.84±0.12 *(5.14±0.16)* | 7.1±0.3 *(6.4±0.16)* | 118±8 |
| naphthalene$^+$ | D$_2$ | 8.2±0.3 *(6.6±0.6)* | 23±1 *(18.8±0.7)* | 190±5 |
| anthracene$^+$ | D$_2$ | 10.2±0.5 *(7.7±0.14)* | 10.1±0.2 *(9.8±0.2)* | c |
| perylene$^+$ | D$_5$ | 18.8±0.2 | 19.3±0.2 | d |

a) life time for the slow component of the TG recovery kinetics for the perprotio D$_0$ state in boric acid; in italics - the life time for a perdeuterio species; the error limits indicate the standard deviation;
b) life time for the fast component (same for perprotio and perdeuterio species);
c) the fast component was not time resolved (this category also includes pyrene-h$_{10}^+$);
d) the fast component is absent.



**FIGURE CAPTIONS.**

Fig. 1

Transient grating signal for the recovery of the ground $D_0$ state of the radical cation observed in a 680 nm pump - 680 nm probe ($D_0 \rightarrow D_2$ band, 0-0 transition) photoexcitation of naphthalene-$h_8^+$ (trace (i), filled circles) and naphthalene-$d_8^+$ (trace (ii), crosses) in a boric acid glass at 295 K. The optical density of this 300 μm thick glass sample was ca. 1. Dashed line (trace (iii), filled diamonds) is a nonresonant optical Kerr effect signal from a thin Suprasil plate, which is taken as an auto correlation trace for the excitation/probe pulse (in this case, a Gaussian pulse of 63 fs fwhm). A solid line (trace (i)) is the least-squares fit obtained by convoluting trace (ii) with a biexponential function (the time constants are given in Table 2). The fast component ( < 1 ps) is from electronic deactivation of the $D_1$ state and the slow component (1-20 ps, ca. 10% of the fast component in weight) is from the vibrational relaxation of a hot $D_0$ state. Trace (ii) is shifted upwards, for clarity. The solid line drawn through the crosses is the same curve as in trace (i), i.e., these is no H/D isotope effect (see also Fig. 3S(a) in the Supporting Information).

Fig. 2

(a) Temperature dependencies of the time constant $\tau_s$ of the slow component for naphthalene-$d_8^+$ (open circles) and perylene-$h_{12}^+$ (open squares). Two data points (15 and 295 K) for naphthalene-$h_8^+$ (open diamonds) are also shown. The error bars give the standard deviation. The temperature dependence (70 to 300 K) of $\tau_s$ for naphthalene-$d_8^+$ can be fit using an empirical formula, $1/\tau_s = k_0 + k_1 \exp(-E_a/RT)$, where $1/k_0$=19.7 ps, $1/k_1$=3.5 ps, and the activation energy $E_a$=31±8 meV. The dashed section is a low-temperature extrapolation. (b) The slow decay kinetics for TG signal in 680 nm pump - 680 nm probe excitation of naphthalene-$d_8^+$ in boric acid glass. Both of these pulses were 160 fs fwhm. The sample temperature is 295 K (filled circles), 190 K (crosses), 150 K (open triangles), 110 K (open squares), and 70 K (open circles). The lines drawn through the symbols are the least-squares exponential fits. The TG kinetics were normalized at the signal maximum (taken as unity). The initial "spike" (see Fig. 1, traces (i) and (ii)) is not shown. As the temperature decreases, the relative weight of the slow component increases and its decay kinetics become slower.

Fig. 3

A scheme of photophysical processes that contribute to the TG signal from a recovering ground $D_0$ state of an aromatic radical cation in a rigid matrix. Photoexcitation of the cation yields an excited $D_n$ state that in < 200 fs relaxes by internal conversion (IC) to a hot $D_0$ state; this process gives rise to the observed subpicosecond kinetics (fast component) that accounts for 70-95% of the total signal. Vibrational relaxation of the hot $D_0$ state, by vibrational energy transfer to the matrix (VET), occurs on the picosecond



time and gives raise to the observed slow kinetics (a "tail" in Fig. 1, trace (i)). These slow kinetics (unlike the rapid internal conversion) are temperature and isotope dependent. The heat transferred to the first solvation shell around the radical cation then slowly dissipates to the matrix bulk. The resulting density grating is not observed (< 1 ns), due to polarization geometry in our experiment.



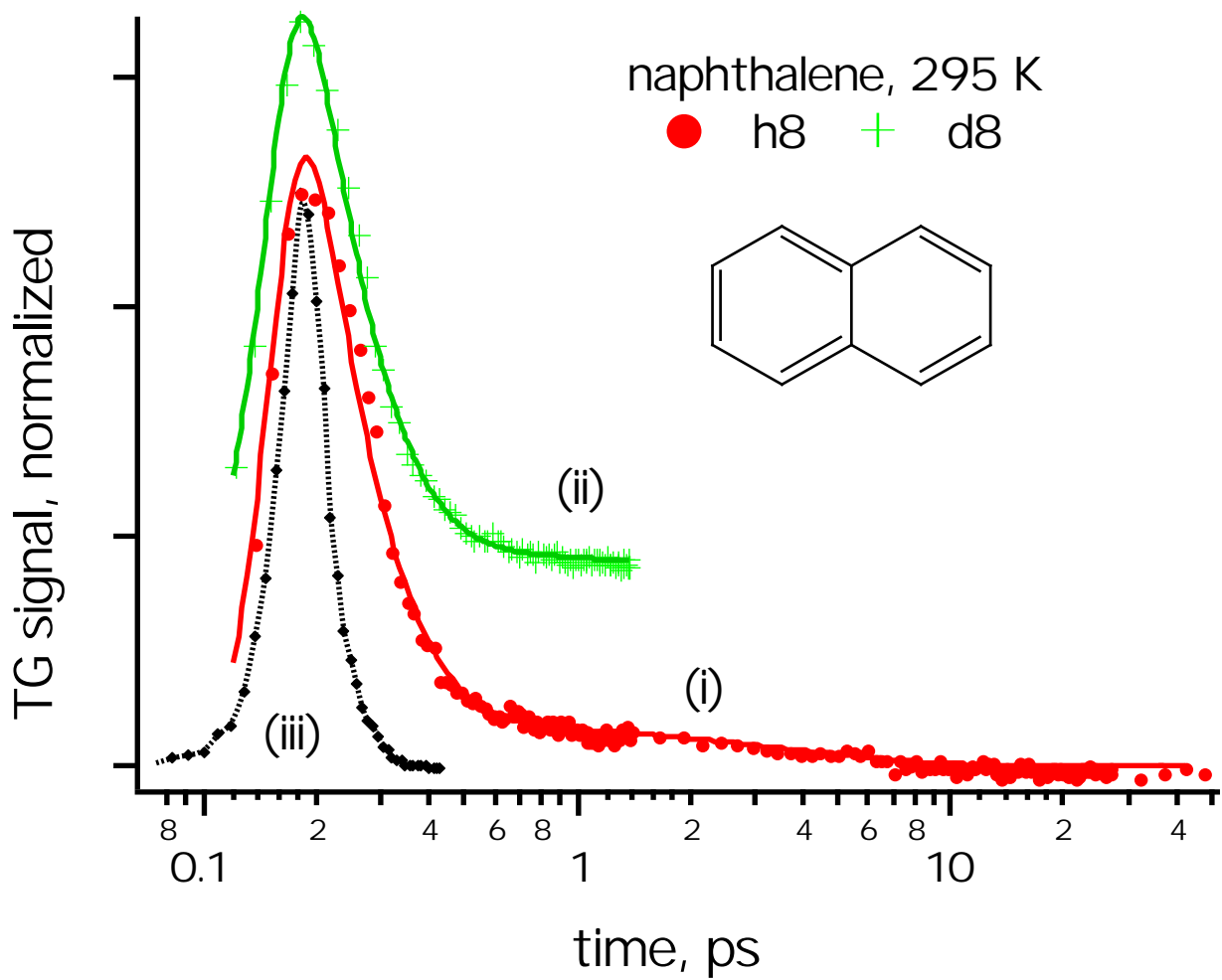

Fig. 1 Crowell et al.

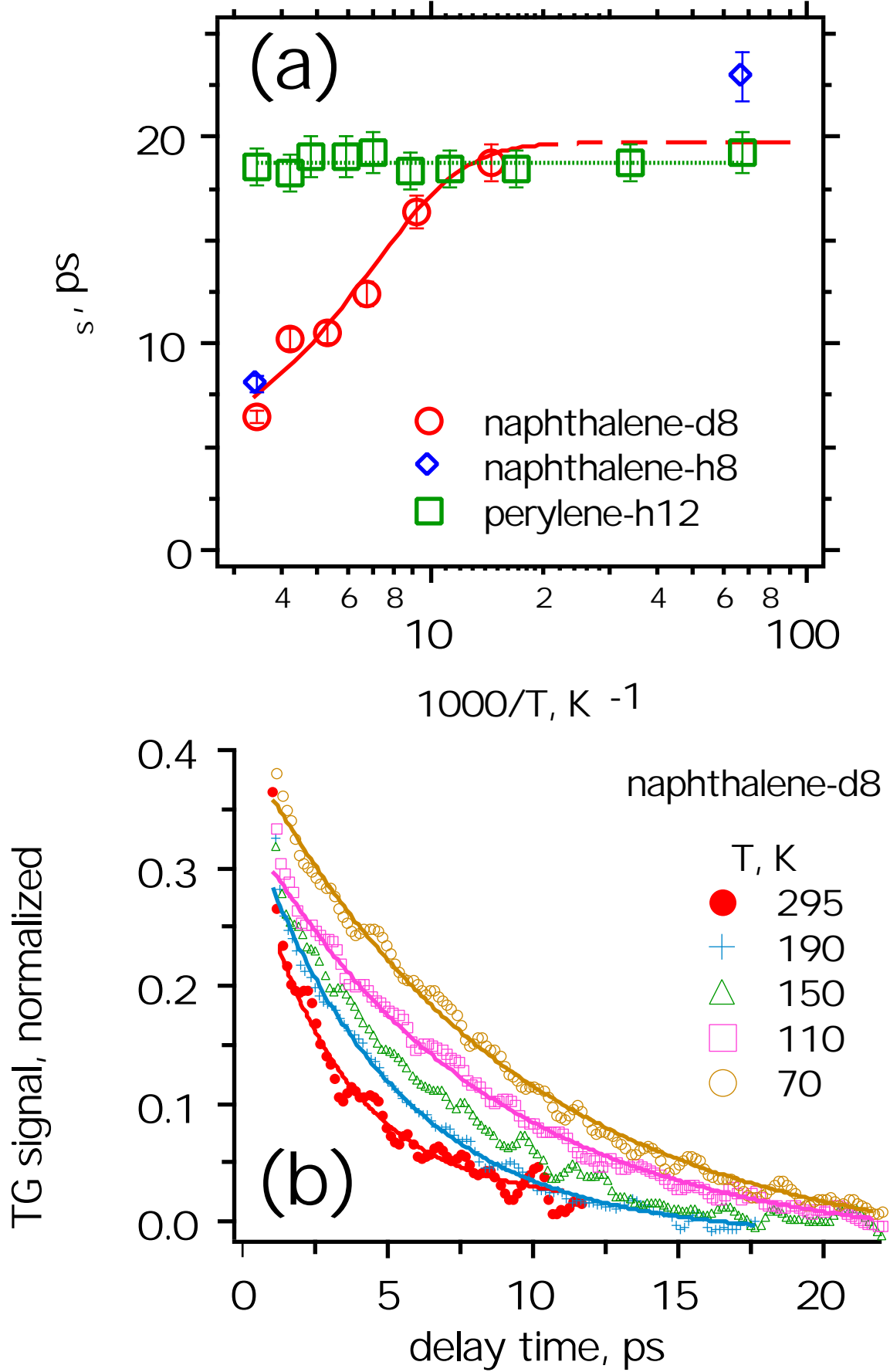

Fig. 2 Crowell et al.

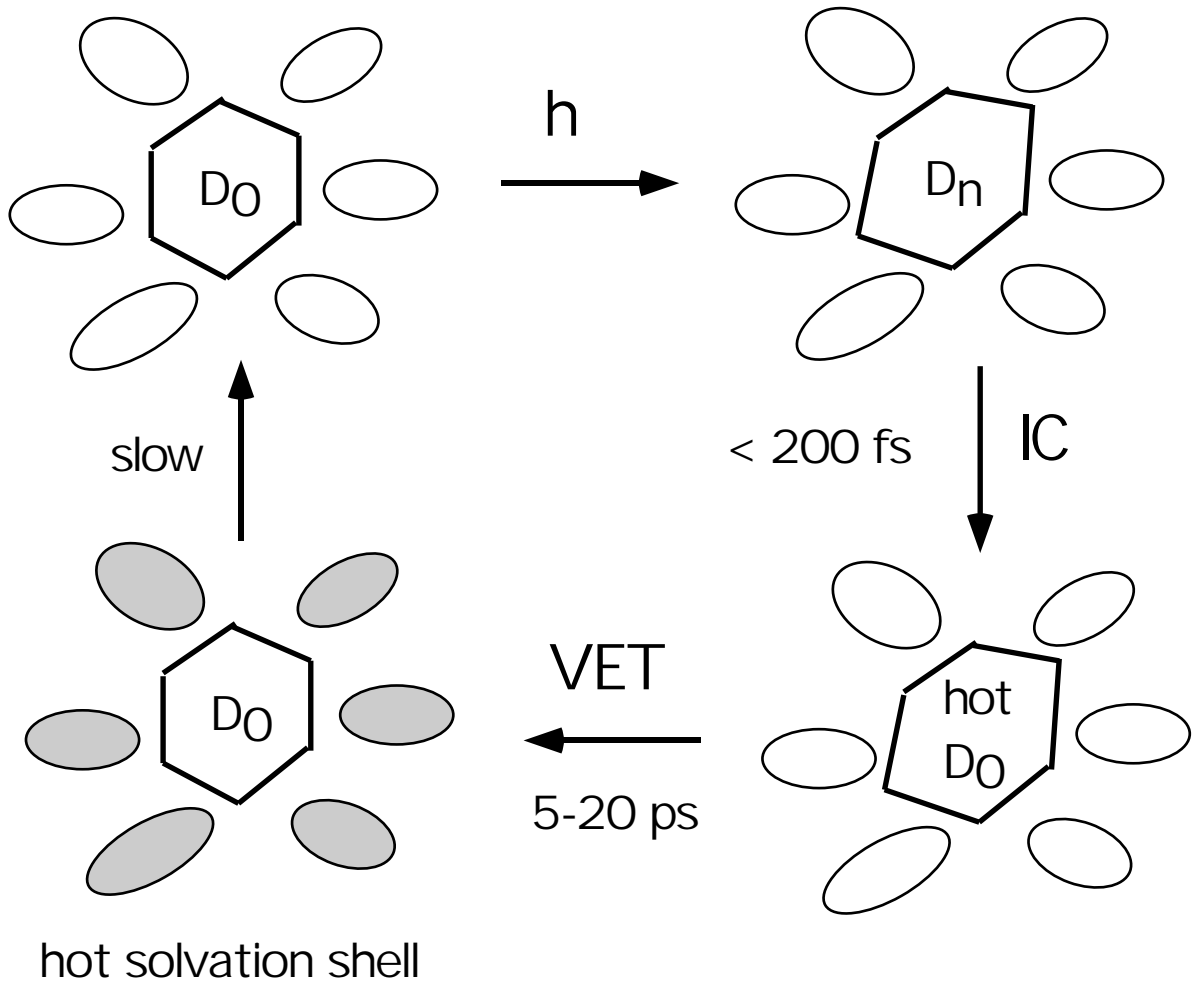

Fig. 3 Crowell et al.



**SUPPLEMENTARY MATERIAL**  JP021832H

*Journal of Physical Chemistry A, Received August 7, 2002*

**Supporting Information.**

**(1S.) Appendix 1: two-step electronic relaxation scenario rebuffed.**

In this section, we examine a scenario in which the fast and the slow decay components in the recovery kinetics of the $D_0$ state are attributed to nonradiative $D_n \rightarrow D_1$ and $D_1 \rightarrow D_0$ transitions, respectively. Since, for obvious reasons, the fast $D_n \rightarrow D_1$ component cannot be observed by following the repopulation kinetics of the *$D_0$ state*, for this scheme to work one needs to postulate the involvement of a light-absorbing $D_1$ state.

The first excited doublet state of naphthalene$^+$ ($D_1$) has ca. 0.89 eV higher energy than the ground $D_0$ state and radiative $D_0 \rightarrow D_1$ transition is symmetry forbidden (the symmetry representations and energetics for *$D_{2h}$* symmetric PAH cations are given in Table 1 and Fig. 2S). The $D_2$ state (the final state upon 1.82 eV excitation of the ground state cation) is ca. 0.92 eV more energetic than the $D_1$ state. The first allowed optical transition of the $D_2$ state is to the $D_5$ state (which is 1.94 eV higher in energy), so it cannot contribute to the TG signal obtained using 1.82 eV probe light. By contrast, a $D_1 \rightarrow D_3$ transition is both energetically and symmetry allowed. A similar situation occurs for anthracene$^+$ and pyrene$^+$ excited to their $D_2$ states (Fig. 2S) and biphenyl$^+$ excited to the $D_3$ state (Fig. 8S). Note that for all of these radical cations the transition dipole moments of the $D_0 \rightarrow D_1$ and $D_1 \rightarrow D_n$ transitions are perpendicular (Fig. 2S and 8S). Since the pump and probe pulses have the same polarization, on average only 1/3 of the generated $D_1$ state cations can be observed in our experiment (assuming no reorientation of the cations in the glass matrix). [1S] Though more than one species can contribute to the transient grating signal, provided that the observed fast kinetics are from a rapid, nonradiative $D_1 \rightarrow D_0$ transition (as argued below), the TG signal would reflect the recovery dynamics of the $D_0$ state even if both the $D_0$ state and the $D_1$ state absorb the probe light.

If the slow component is due to $D_1 \rightarrow D_0$ transition, one may expect that the corresponding rate constants would follow the energy gap law (see the Discussion). In accord with these expectations, the $\tau_s$ times for the picosecond component at 295 K systematically increase with the $D_1$-$D_0$ energy gap, from biphenyl$^+$ to perylene$^+$ (though this correlation vanishes when low-temperature lifetimes are compared). For pyrene$^+$ and anthracene$^+$ ($D_0 \rightarrow D_2$ excitation), the $D_2$-$D_1$ and $D_1$-$D_0$ gaps are 3,000 and 10,700 cm$^{-1}$ and 5,000 and 9,400 cm$^{-1}$, respectively. [16] Conceivably, the rapid relaxation observed for these two cations ($\tau_f < 50$ fs) could be from a $D_2 \rightarrow D_1$ transition and the picosecond



component from a slower $D_1 \rightarrow D_0$ transition. For anthracene$^+$ and naphthalene$^+$ in solid Ne and Ar, [16] the line widths of the *0-0* lines are 120 and 50 cm$^{-1}$, respectively (the $D_0 \rightarrow D_2$ transition). If these line widths are due to lifetime broadening, one obtains estimates of 44 and 106 fs, respectively. These lifetimes compare favorably with the experimental lifetimes for the fast component. For biphenyl$^+$ ($D_0 \rightarrow D_3$ excitation), [17] there are two nearly degenerate states, $D_1$ and $D_2$, midway between the $D_0$ and $D_3$ states (Fig. 8S and the caption to this figure). The $D_3$-$D_{1,2}$ gap is 6,300 cm$^{-1}$, which is larger than the $D_2$-$D_1$ gap in pyrene$^+$ and anthracene$^+$ and smaller than the $D_2$-$D_1$ gap in naphthalene$^+$ (7,400 cm$^{-1}$). [16,17] Following this trend, the fast component for biphenyl$^+$ decays slower than the fast component for anthracene$^+$ and faster than the fast component for naphthalene$^+$. For perylene$^+$ ($D_0 \rightarrow D_5$ excitation) [16] and tetracene$^+$ ($D_0 \rightarrow D_1$ excitation), [18] the $D_0$-$D_1$ gaps are large (> 11,000 cm$^{-1}$) which accounts for their slow decay kinetics. The lack of a fast component for perylene$^+$ can be accounted for by high density of the $D_n$ states (the largest gap of 4600 cm$^{-1}$ is for the $D_5 \rightarrow D_4$ transition), [16] i.e., by extremely rapid $D_n \rightarrow D_1$ relaxation.

Despite some correlation between the $\tau_f$ and $\tau_s$ times and the corresponding energy gaps, the scenario in which the fast/slow components arise from the $D_n \rightarrow D_1$ and $D_1 \rightarrow D_0$ transitions, respectively, is not supported by our kinetic data:

First, the energetics argument is not consistent. For naphthalene$^+$, the $D_2$-$D_1$ and $D_1$-$D_0$ gaps are comparable, yet one component is 40 times slower than another (Fig. 1 and Table 2). A similar discrepancy is observed for biphenyl$^+$, for which the $D_3$-$D_{1,2}$ and $D_{1,2}$-$D_0$ energy gaps are also comparable (Fig. 8S). It is known that the H/D isotope effect on the rate of nonradiative transitions (e.g., $T_1 \rightarrow S_0$ transitions) inversely correlates with the energy gap, as only high-frequency C-H stretching modes contribute to this effect. [10] For energy gaps smaller than 15,000 cm$^{-1}$, the isotope effect is usually quite small. [10] For naphthalene$^+$, anthracene$^+$, and biphenyl$^+$, the $D_0$-$D_1$ gaps are smaller than 10,000 cm$^{-1}$, yet the corresponding $\tau_s$ times change considerably with H/D substitution. If the picosecond components were always due to $D_1 \rightarrow D_0$ transitions, why would the time constants $\tau_s$ of these transitions vary with temperature and isotope composition for some PAH cations (such as naphthalene, biphenyl, and anthracene) but not others (such as perylene) and why would the fast components not exhibit such variations?

Second, simple kinetic considerations suggest that if the $D_1$ state contributed to the TG kinetics, larger variations in the time profiles with respect to the absorption properties of the $D_0$ and $D_1$ states should be observed. Let $K=\tau_f/\tau_s$ be the ratio of the fast ($D_n \rightarrow D_1$) and slow ($D_1 \rightarrow D_0$) relaxation times and $\rho=\varepsilon(D_1)/3\varepsilon(D_0)$ be the ratio of molar extinction coefficients $\varepsilon$ of the $D_1$ and $D_0$ states corrected by the polarization factor of 1/3. Then, the transient grating signal is proportional to



$$S(t) \propto \{(1-\rho)/(1-K) \exp(-t/\tau_s) + (K-\rho)/(1-K) \exp(-t/\tau_f)\}^2 \quad (1S)$$

convoluted with the time profile of the excitation pulse. Since $K \ll 1$, eq. (1S) gives essentially bimodal kinetics with a fast and slow component. Using eq. (1S), it is possible to fit the observed TG kinetics for naphthalene$^+$ and biphenyl$^+$ shown in Figs. 1, 3S(a), and 4S(b), with $\tau_f$ of 140±7 fs and 76±7 fs, respectively (and the same $\tau_s$ values as given in Table 2) and $\rho$ of 0.79 and 0.76, respectively. Parameter $\rho$ being close to unity helps to explain the kinetic data: for $K \ll 1$, the relative weight of the slow component is given by $(1-\rho)^2$. In order to obtain the observed range of 2-10% for these weights, $\rho$ should always be 0.7-to-0.9. This does not seem plausible: there is no reason to expect that $\varepsilon(D_1)$ is always approximately three times greater than $\varepsilon(D_0)$. Moreover, as the weight of the slow component is a function of $\rho$ only (for $K \ll 1$), it is difficult (within this model) to account for the strong variations of the weight of the slow component with temperature and H/D substitution that were observed in our experiments (see Figs. 2(b), 5S, and 6S).

Third, if the slow component was due to the $D_1 \rightarrow D_0$ conversion, it is surprising that no vibrational relaxation of the resulting "hot" $D_0$ state is observed afterwards. [20-22] The excess energy of a "hot" $D_0$ state is greater than 10,000 cm$^{-1}$. In such a case, the line shapes of the $D_0 \rightarrow D_n$ band for the "hot" and thermalized $D_0$ states have to be different. [19] Specifically, the spectrum of the "hot" state should be broader [19a] and, possibly, red-shifted. [21] Given that the excitation bandwidth of the probe pulse, ca. 220 cm$^{-1}$, is less than the homogeneous line width of a PAH cation trapped in boric acid glass, 300-500 cm$^{-1}$, [11,18b] even a small difference between the spectral lines for the "hot" and the relaxed $D_0$ states should yield a measurable TG signal. There should be a TG component that corresponds to the vibrational relaxation of the "hot" $D_0$ state. We argue that the slow component corresponds to this vibrational relaxation dynamics.

**(2S.) Appendix 2: intermolecular VET in boric acid glass.**

Below, we speculate on the possible mechanisms for isotope- and temperature-sensitive vibrational energy transfer from the "hot" D state of a planar PAH cation to the glass matrix which are postulated in the "two-step" relaxation model.

The H/D isotope effect may originate through a VET that involves low-frequency C-H modes of the PAH cation [25] and vibrational modes of glass-forming superstructural units in the first solvation shell. In vitreous (v-) $B_2O_3$, 85% of boron atoms are incorporated in planar trimeric $B_3O_6$ ("boroxol") rings whose breathing mode at 808 cm$^{-1}$ is the strongest peak in the Raman spectrum of this solid. [2S] A matching peak was observed in the vibrational density of state (VDOS) of the v-$B_2O_3$ that was obtained in inelastic neutron scattering experiments of Sinclair and co-workers. [2S] It is likely that PAH molecules in boric acid are sandwiched between these boroxol rings (a related crystalline solid, $\alpha$-HBO$_2$, consists of infinite sheets of hydrogen-bonded boroxol rings), [3S] i.e., these rings are the most likely acceptor of the vibrational energy from a "hot" $D_0$ state of a PAH cation. The 808 cm$^{-1}$ breathing mode of the boroxol ring is close in energy to out-of-plane C-H modes of planar PAH cations (that cover the interval between 770 to 900 cm$^{-1}$ and make a large contribution to the calculated VDOS and IR emission spectra of the PAH cations). [25] For example, naphthalene-h$_8^+$ has one of these modes at 849 cm$^{-1}$, anthracene-h$_{10}^+$ - at 912 cm$^{-1}$, and pyrene-h$_{10}^+$ - at 861 cm$^{-1}$. [25] The close match in energy between the two most prominent VDOS peaks for the cation and the matrix would account for the efficient VET from the out-of-plane C-H modes of the PAH cation to a nearby boroxol ring. Typical isotope effects on the vibrational frequency for these C-H modes in the perdeuterio and perprotio cations are 1.15-1.2 (as compared to 1.33-1.36 for the high-frequency C-H stretch modes). [25] Thus, the corresponding VET rate could exhibit an isotope effect. Since the effective temperature of the "hot" state is much higher than that of the matrix, this transfer rate does not depend on the sample temperature.

By contrast, cooling of the first solvation shell depends on the sample temperature whereas it obviously does not depend on the isotope composition of the dopant. We speculate that the heat transfer from the first solvation shell to the glass bulk occurs mainly by emission of acoustic phonons. [4S] For T>10 K, the typical phonon mean free path in a glass is 1-3 nm, [4S] which accounts for efficient transfer of heat from the "hot" first solvation shell to the glass matrix. The transfer rate should approximately correlate with the thermal conductivity of the glass. For vitreous $B_2O_3$ (whose structure is similar to that of boric acid glass) the macroscopic thermal conductivities at 50 and 300 K are 0.2 and 0.52 W/(m·K), respectively. [23] As seen from Fig. 2(a), $1/\tau_s$ for naphthalene-d$_8^+$ changes by approximately the same factor. For all amorphous solids, the thermal conductivity is nearly constant between 2 and 100 K (the so-called "plateau" region); above 100 K, the conductivity steadily increases with the temperature. [23,4S] Following the same trend, $\tau_s$ changes little below 100 K and systematically decreases between 100 and 300 K (Fig. 2(a)).

**(3S.) Additional references.**

1S. The statistical factor of 1/3 is due to the fact that the magnitude of a population grating is proportional to $\cos^2\theta$, where $\theta$ is the angle between the transition dipole for a ground state cation and the direction of electric field in a pump pulse. The polarizations of the pump, probe, and scattered beams all point in the same direction. Therefore, when the transition dipole of the probed state (e.g., $D_1$ state) has the same orientation as that of the ground $D_0$ state, the overall weight factor is $\cos^2\theta \times \cos^2\theta$. When the transition dipole of the probed state (e.g., $D_1$ state) is perpendicular to that of the $D_0$ state, this factor is $\cos^2\theta \times \sin^2\theta \times \cos^2\chi$, where $\chi$ is the asimuthal angle of the $D_1$ state dipole in a plane that is normal to the $D_0$ state dipole. Angle averaging of these two factors gives 2/5 and 2/15, respectively, and the ratio is 1/3.

2S. Hannon, A. C.; Grimley, D. I.; Hulme, R. A.; Wright, A. C.; Sinclair, R. G. *J. Non-Cryst. Solids* **1994**, *177*, 299 *and references therein.*

3S. V. F. Ross, J. O. Edwards, in *The Chemistry of Boron and Its Compounds,* edited by E. L. Muetterties (Wiley, New York, 1967), p. 155; W. H. Zachariasen, Acta Cryst. **1954**, *7*, 305.

4S. Zeller, R. C.; Pohl, R. O. *Phys. Rev. B* **1971**, *4*, 2029; Stephens, R. B. *Phys. Rev. B* 1972

**(4S.) Table 1S.**

Quantum yields for 532 nm photon induced "hole injection" in the room-temperature *trans*-decalin and 2-propanol.

| radical cation | $IP_{gas}$, eV [a] | $\phi_1$ [b] | $\phi_2$ [c] |
|---|---|---|---|
| naphthalene$^+$ | 8.14 | 0.11 | 0.46 |
| biphenyl$^+$ | 7.95 | 0.046 | 0.30 |
| pyrene$^+$ | 7.41 | - | 0.073 |
| perylene$^+$ | 6.9 | 0.0065 | 0.022 |

a) gas-phase first ionization potential of the parent molecule (ref. 26); for anthracene and biphenylene, the ionization potentials are 7.45 and 7.95 eV, respectively.
b) quantum yield of reaction (1) in *trans*-decalin (ref. 26).
d) quantum yield of reaction (2) in 2-propanol (ref. 27).

**(5S.) Captions to figures 1S to 8S.**

Fig. 1S

The structures of parent aromatic molecules discussed in the paper: (a) naphthalene, (b) anthracene, (c) pyrene, (d) tetracene, (e) biphenyl, and (f) perylene. All molecules and radical cations with exception of biphenyl have $D_{2h}$ symmetry. Biphenyl has $D_2$ symmetry; *ab initio* calculations of Heidenreich et al. (ref. 17e) suggest that the dihedral angle between the two phenyl rings in biphenyl$^+$ is 30-40º as compared to 40-50º in the neutral matrix-isolated molecule.

Fig. 2S

State diagrams for $D_{2h}$ symmetric planar radical cations *(from left to right)* of naphthalene (Np), anthracene (An), pyrene (Py), and perylene (Pe). The results of *ab initio* calculations of Negri and Zgierski (ref. 16) are used. The symmetry representations of the $D_n$ states are indicated in the diagram. Only doublet states with excess energy < 3.5 eV are shown. Bold arrows indicate the excitation bands pumped and probed in our TG experiments. Thin arrows indicate symmetry-allowed transitions of the $D_n$ states (at the probe energy) that can contribute to the TG signal. Thin lines indicate some relevant symmetry allowed transitions in the doublet manifold. Red (dashed lines) is for optical transitions in which the transient electric dipole is in the direction of the *y* axis (Fig. 1S); green (solid line) is the same for the *x* axis.

Fig. 3S

(a) A comparison between the fast TG kinetics for naphthalene-$h_8^+$ *(filled circles)* and naphthalene-$d_8^+$ (two different runs are shown by empty squares and triangles) in boric acid glass at 295 K. $D_0 \to D_2$ band, 0-0 transition photoexcitation. The autocorrelation trace for a 70 fs fwhm, 680 nm pulse (both pump and probe) is indicated by the solid line. Within experimental error, there is no isotope effect on the subpicosecond component. (b) A double logarithmic plot for the decay kinetics of TG signal for naphthalene-$h_8^+$ in boric acid glass at 15 K. The radical cation was pumped and probed using 680 nm, 160 fs pulses. Note the increased weight for the slow ($\tau_s \approx 23$ ps) component at the low temperature. The dotted line is the autocorrelation trace of the excitation pulse. Open and closed circles are for the $\Delta t=20$ fs and $\Delta t=150$ fs scans, respectively. The solid line drawn through the picosecond kinetic trace is a single-exponential least-squares fit.

Fig. 4S

Subpicosecond kinetics for the TG signal from (a) $h_{10}$ and $d_{10}$ isotopomers of anthracene$^+$ (filled circles and open squares, respectively) and (b) $h_{10}$ and $d_{10}$ isotopomers of biphenyl$^+$ (vertical bars, see the legend in the plot) in room-temperature boric acid glass. The pump and probe pulses are 70 fs fwhm, 720 nm for anthracene$^+$ ($D_0 \to D_5$ excitation) and 680 nm for biphenyl$^+$ ($D_0 \to D_3$ excitation). The pulse autocorrelation traces are shown by filled diamonds (a) and black bars (b); Gaussian curves are drawn through these traces. The vertical error bars indicate 95% confidence limits obtained using a bi-tail Student's t-distribution for an average of 100 scans. For anthracene$^+$, the kinetics for $h_{10}$ and $d_{10}$ isotopomers are identical within the experimental error; these kinetics closely follow the time profile of the excitation pulse. For biphenyl$^+$, there is a subpicosecond component. Within the experimental error, the time constant $\tau_f$ is the same for both isotopomers. Note that the slow component for biphenyl-$h_{10}^+$ has higher weight than the slow component for biphenyl-$d_{10}^+$. In more detail, these slow kinetics can be seen in Fig. 6S.

Fig. 5S

(a) Decay kinetics of the TG signal anthracene-$h_{10}^+$ (quasi logarithmic time scan) in boric acid glass at 295 K. These kinetics were obtained using 720 nm, 70 fs pump and probe pulses ($D_0 \to D_2$ excitation); the autocorrelation trace for this pulse is shown by a solid black line. Vertical bars indicate 95% confidence limits. The picosecond "tail" (multiplied by 5 times) is barely seen in this kinetics. (b) Using 160 fs pulses and more extensive averaging, the slow component can be better observed (the initial "spike" is not shown).

The kinetics for anthracene-$d_{10}^+$ and anthracene-$h_{10}^+$ in boric acid glass, at 15 K and 295 K, are plotted together (see the legend in the plot). These kinetics were normalized at the signal maximum. The lines drawn through the symbols are single-exponential least-squares fits of the kinetic traces. The time constants are given in Table 2. The relative weight of the slow component in the perprotio cation is greater than in a perdeuterio cation, both at 15 K and 295 K.

Fig. 6S

Slow decay kinetics of the TG signal for (a) biphenyl-$h_{10}^+$ and (b) biphenyl-$d_{10}^+$ in boric acid glass (filled circles for 295 K and empty squares for 15 K). These kinetics were obtained using 680 nm, 160 fs pump and probe pulses ($D_0 \rightarrow D_3$ excitation). The initial "spike" is not shown. The kinetics were normalized at the signal maximum. (c) A comparison between the kinetics for biphenyl-$h_{10}^+$ (filled circles) and biphenyl-$d_{10}^+$ (empty circles) at 15 K. The lines drawn through the symbols in (a)-(c) are single-exponential least-squares fits of the kinetic traces.

Fig. 7S

Evolution of the decay kinetics of TG signals for perylene-$h_{12}^+$ in boric acid glass as a function of temperature T of the sample (15 to 240 K, see the legend in the plot). These kinetics were obtained using 540 nm, 160 fs fwhm pump and probe pulses ($D_0 \rightarrow D_5$ excitation). An offset 170 K trace is shown separately (crosses); the line through the symbols is a single-exponential least-squares fit. The time constants $\tau_s$ are given in Fig. 2(a), *open squares*. For perylene, the subpicosecond component is lacking, and the slow decay kinetics are temperature independent.

Fig. 8S

Energy diagram for $D_2$ or $D_{2h}$ symmetric biphenyl$^+$. The $D_2$ group representations can be obtained from the $D_{2h}$ group representations by dropping the parity index. On the left side, the energy levels estimated from photoelectron spectroscopy (PES) data of refs. 17a and 17b are shown. Following Heidenreich et al.,[17e] the states are classified for a planar $D_{2h}$ symmetric cation. On the right, absorption spectroscopy data (ABS), by Puiu et al.[17d] for biphenyl-$h_{10}^+$ in solid Ar, are shown. The dashed lines are energy levels for the *1* $A_u$, *3* $B_{3u}$, and *4* $B_{3u}$ levels in a planar radical cation obtained in the LNDO/S PERTCI calculation by Heidenreich et al. (1.53, 2.22, and 3.56 eV);[17e] these energies were reduced by an equal amount to match the theoretical and experimental energies for the $D_0 \rightarrow D_3$ and $D_0 \rightarrow D_4$ transitions. According to these calculations,[17e] the best match for the $D_0 \rightarrow D_3$ transition is obtained in a nonplanar $D_2$ symmetric cation with a dihedral angle of 35º between the phenyl rings (in a solid matrix, there could be a distribution of these dihedral angles). The bold line indicates the $D_0 \rightarrow D_3$ transition at 1.82 eV (the transition dipole moment is along the z axis perpendicular to the line of sight). A thin line indicates a possible (symmetry allowed) $D_2 \rightarrow D_4$ transition at the same probe pulse energy (the transition dipole is along the long y axis of the cation). Dashed red lines indicate some other symmetry allowed transitions in the doublet manifold of the radical cation. Note that the *1* $A_u$ and *2* $B_{1g}$ states are nearly degenerate.

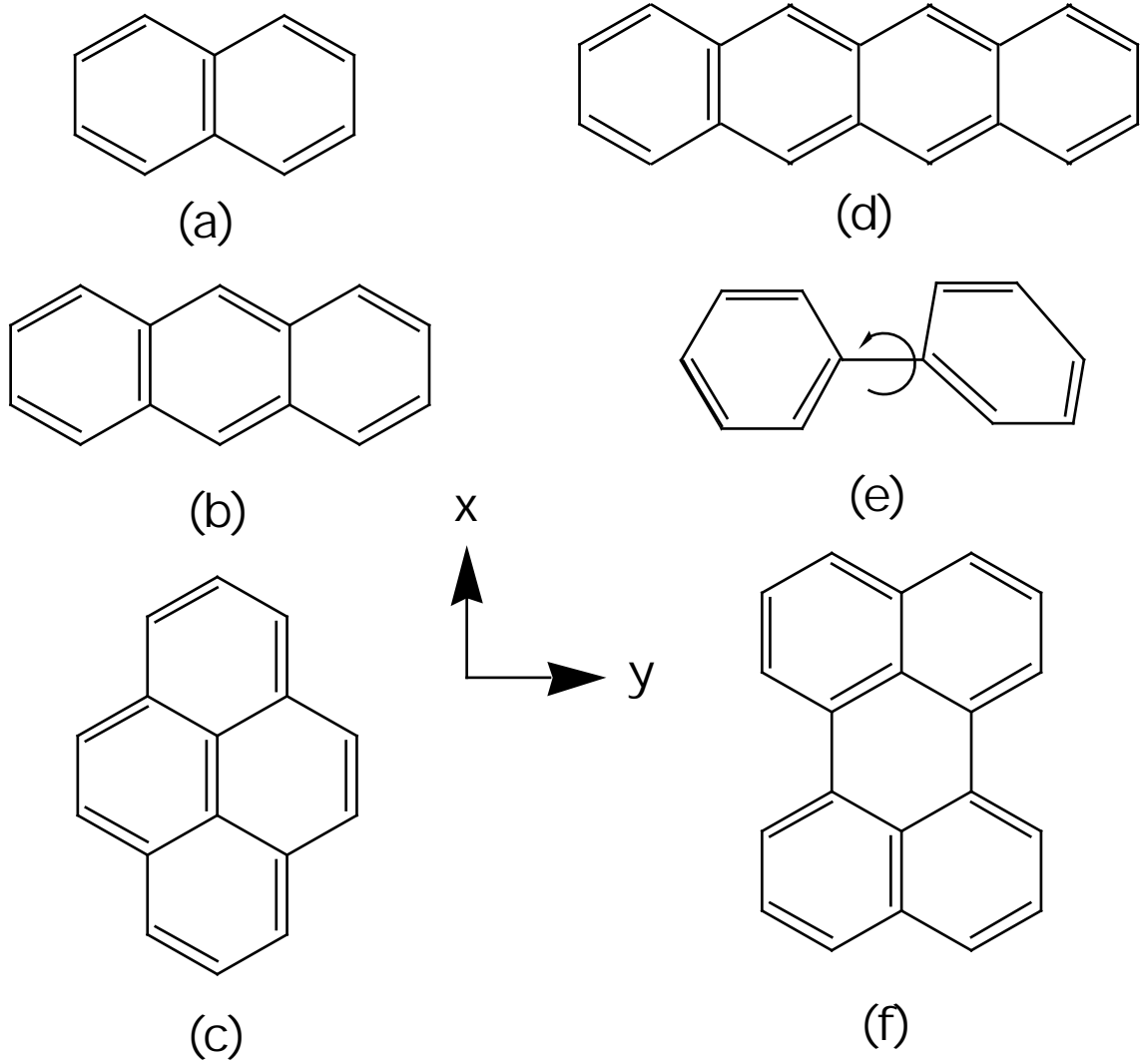

**Fig. 1S Crowell et al.**

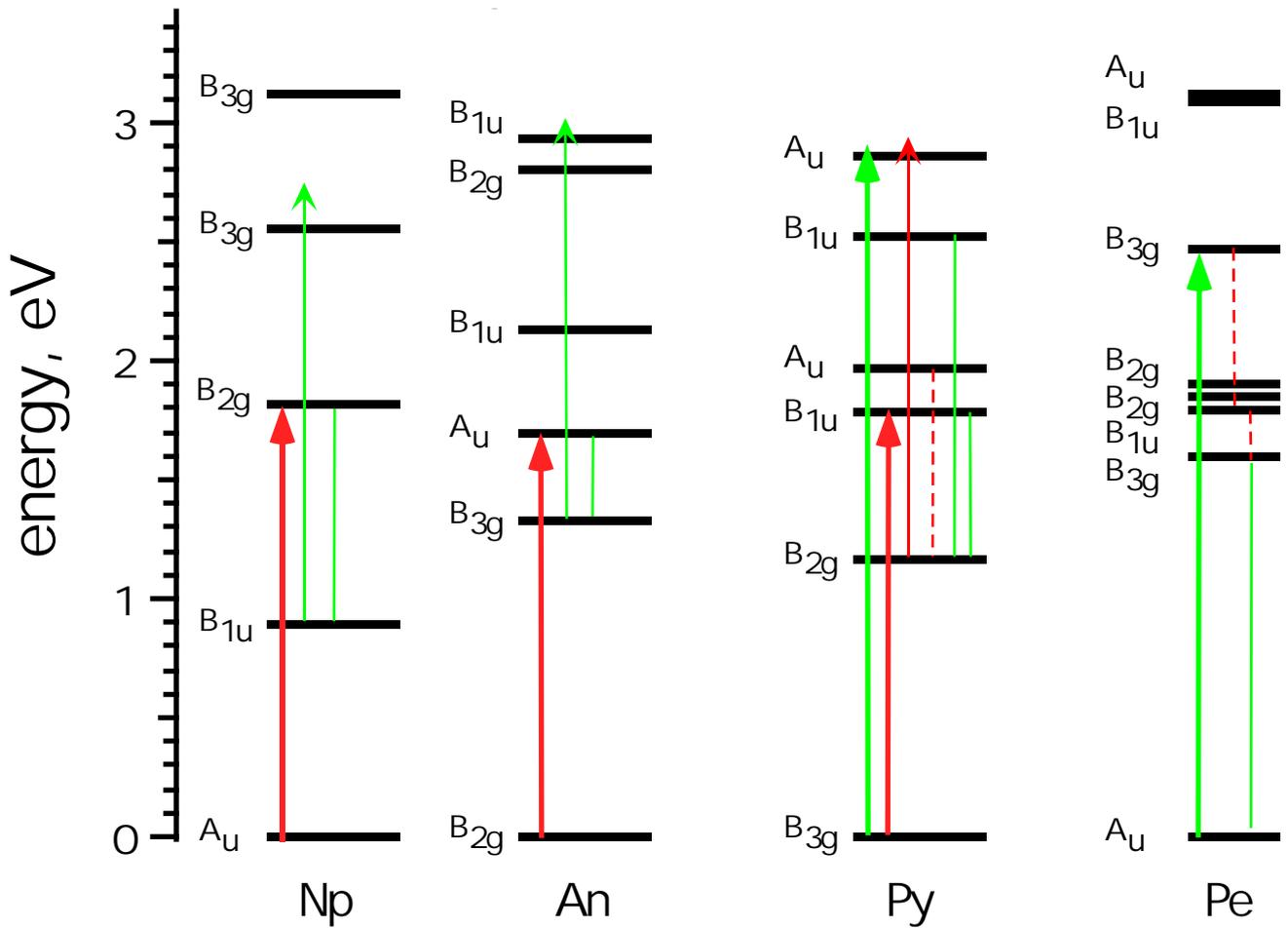

Fig. 2S Crowell et al.

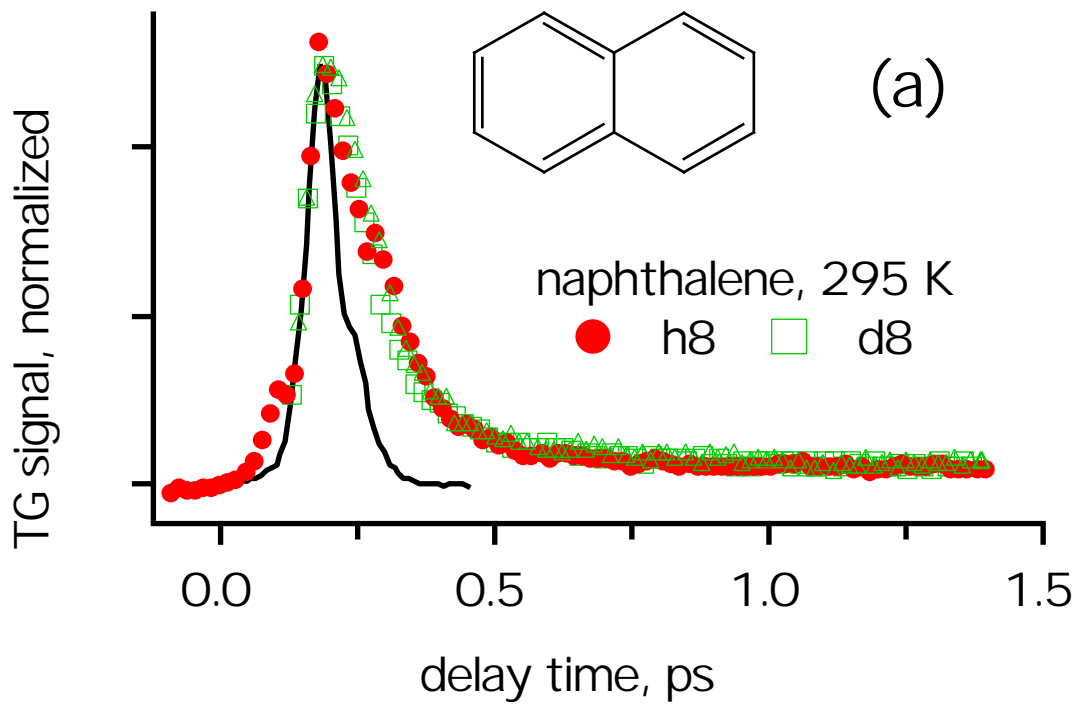
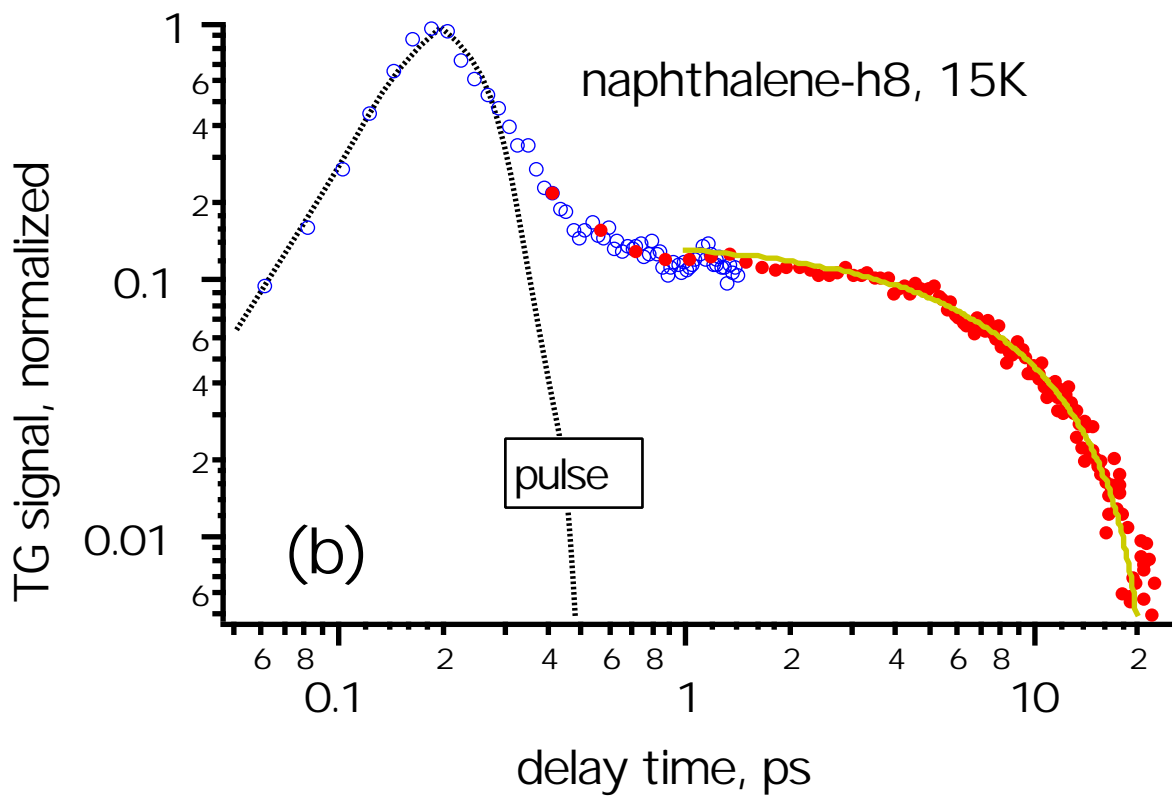

Fig. 3S Crowell et al.

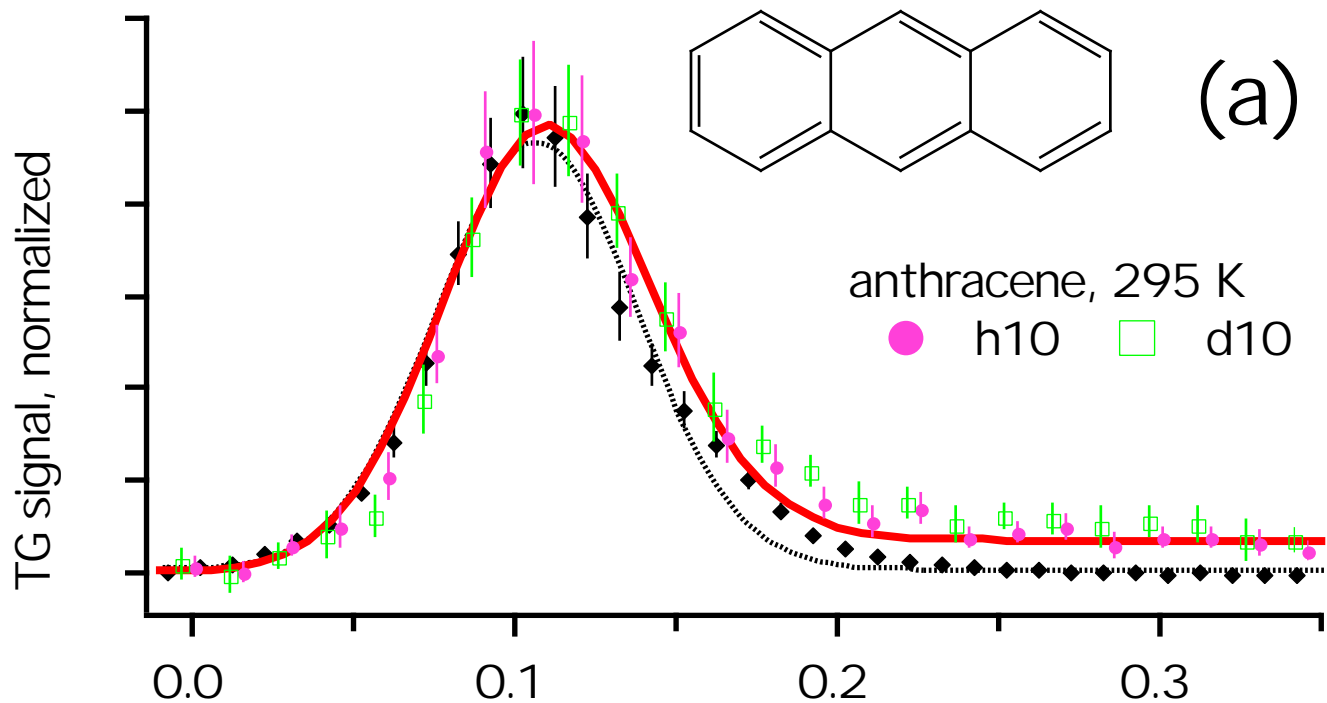
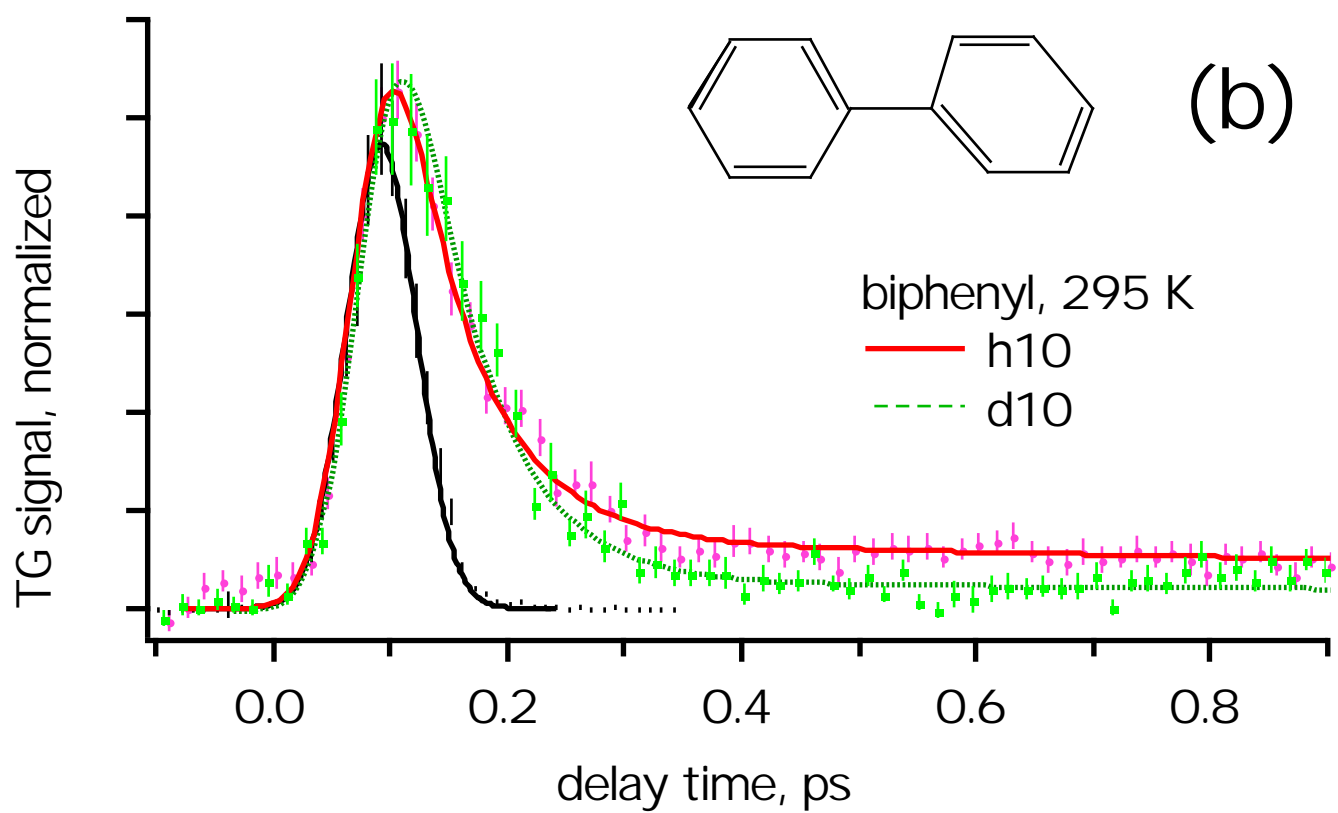

Fig. 4S Crowell et al.

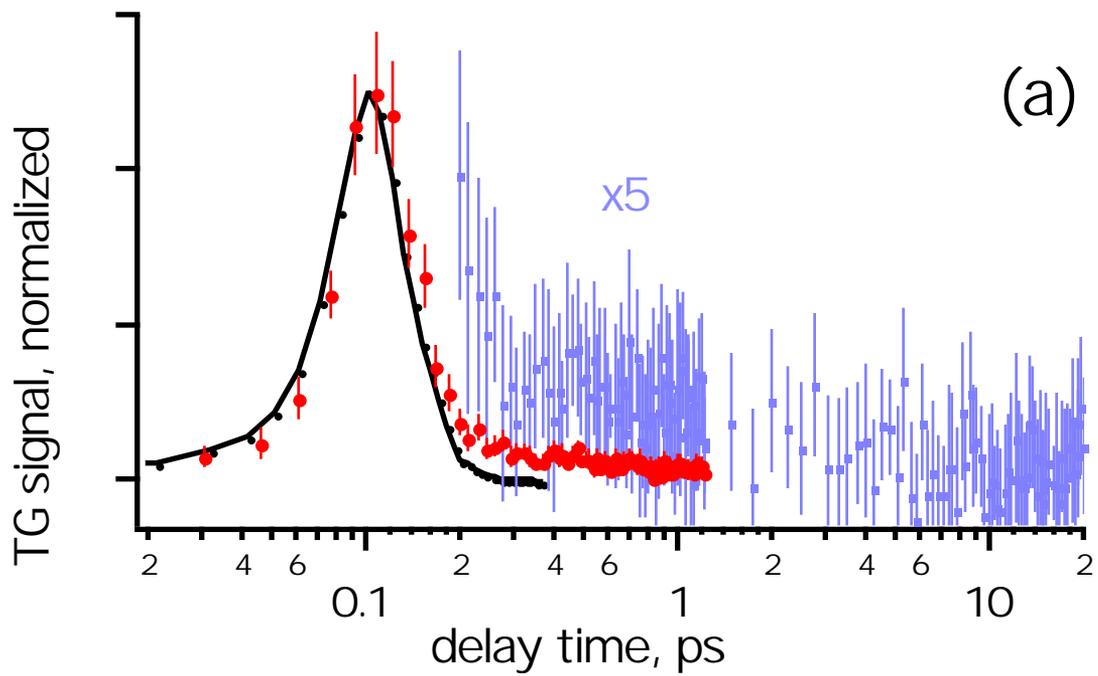
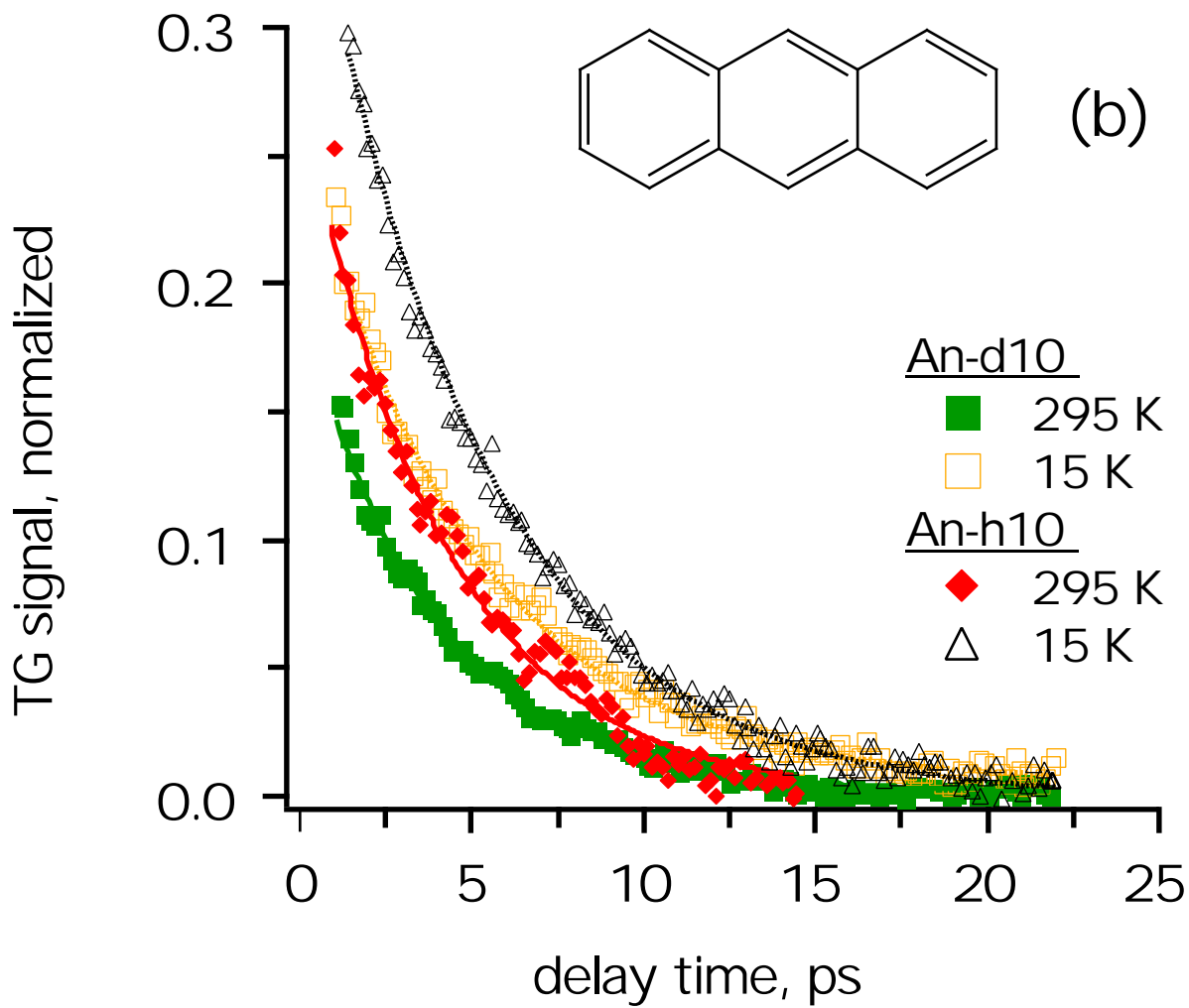

Fig. 5S Crowell et al.

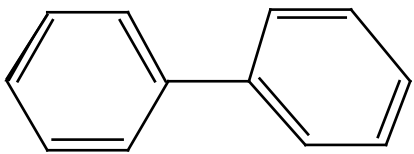
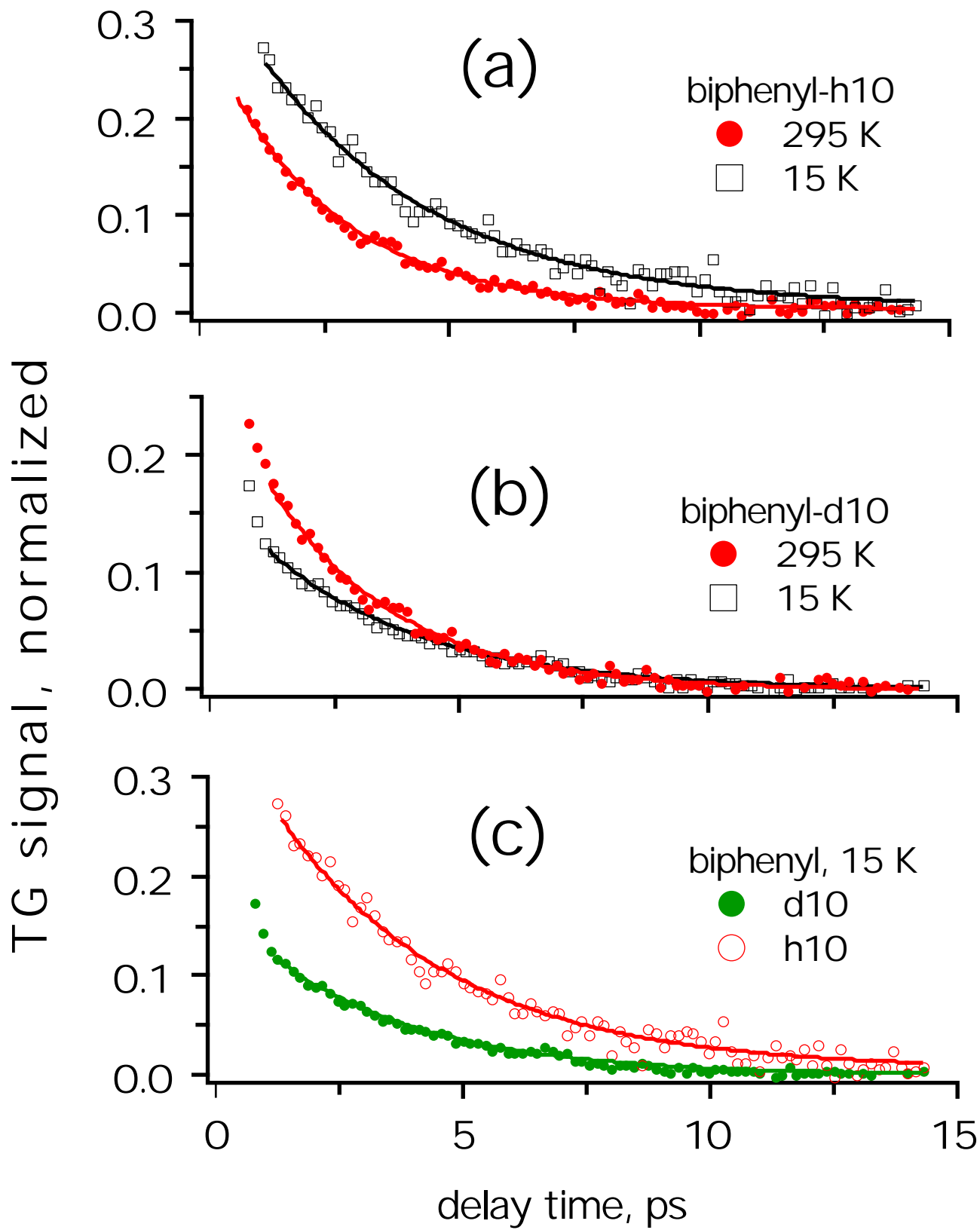

Fig. 6S Crowell et al.

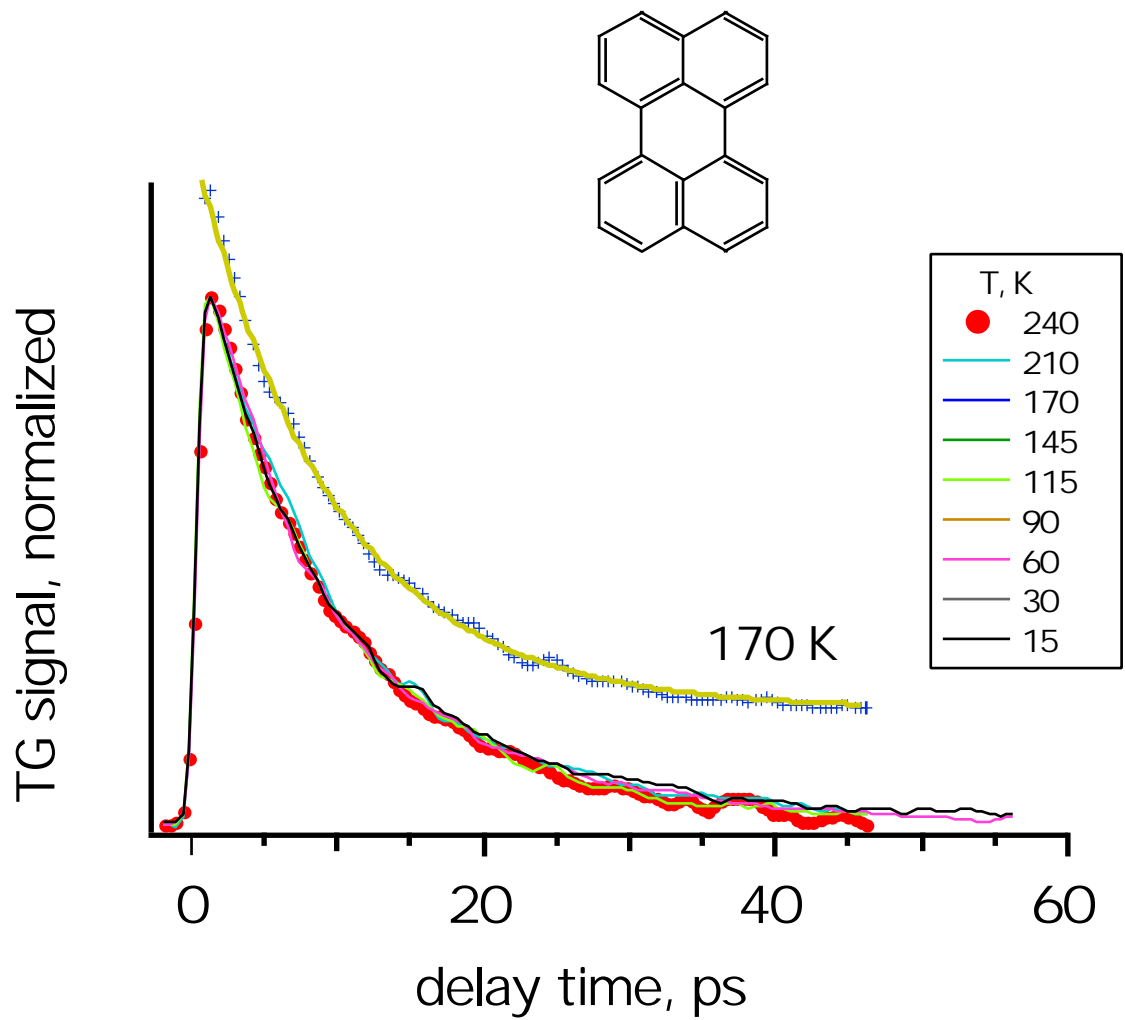

Fig. 7S Crowell et al.

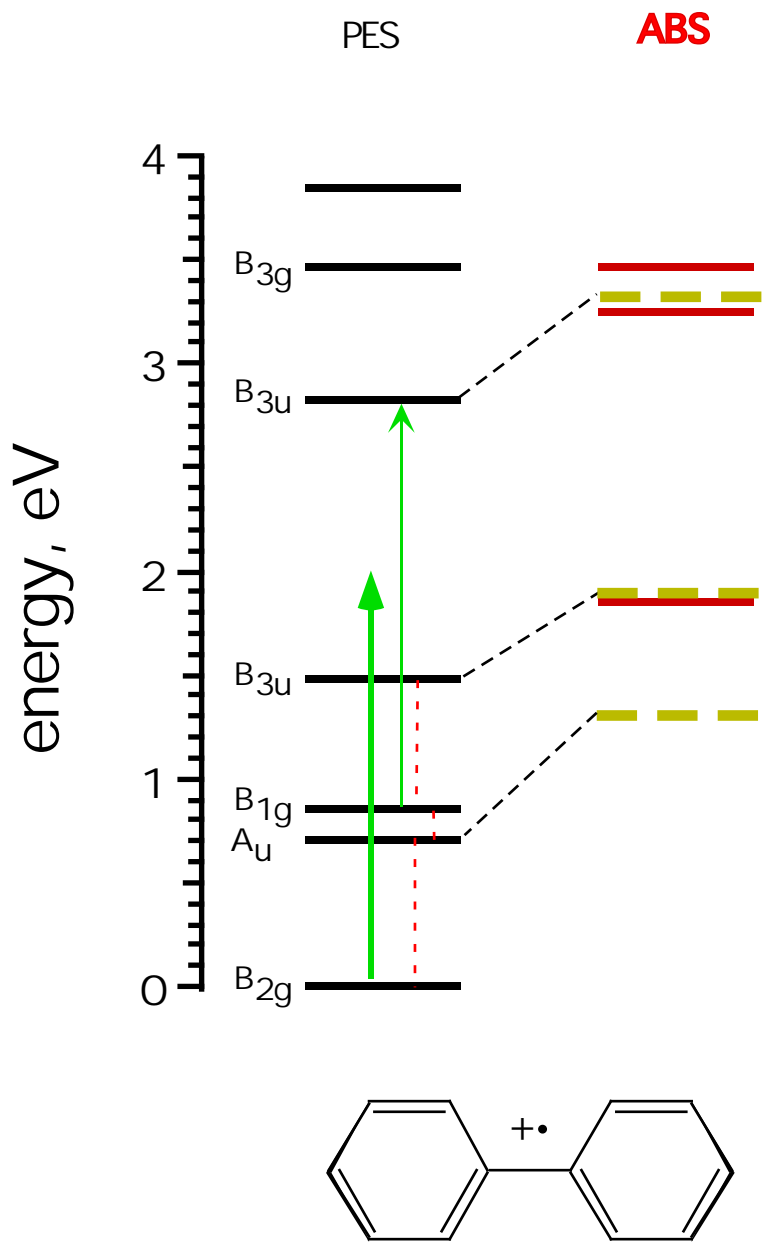

**Fig. 8S Crowell et al.**